\documentclass[3p,authoryear,a4paper]{elsarticle}
\usepackage{graphicx}
\usepackage{xcolor}
\usepackage{dsfont}
\usepackage{amsmath}
\usepackage{amssymb}
\usepackage{amsfonts}
\usepackage{amsbsy}
\usepackage{amsthm}
\usepackage{array}
\usepackage{booktabs}
\usepackage{verbatim}
\usepackage[english]{babel}
\usepackage{float}
\usepackage{subfigure}
\usepackage{epsfig}
\usepackage{multirow}
\usepackage{natbib}
\usepackage{hyperref}
\usepackage{float}
\usepackage[font=small,skip=3pt]{caption}
\usepackage{graphicx}
\usepackage{pdfpages}
\usepackage[toc,page]{appendix}

\newcolumntype{C}[1]{>{\centering\let\newline\\\arraybackslash\hspace{0pt}}m{#1}}
\renewcommand\appendix{\par
	\setcounter{section}{0}%
	\setcounter{subsection}{0}%
	\setcounter{table}{0}
	\setcounter{figure}{0}
	\gdef\thetable{\Alph{table}}
	\gdef\thefigure{\Alph{figure}}
	\gdef\thesection{\Alph{section}}
	\setcounter{section}{0}}

\newcolumntype{L}[1]{>{\raggedright\let\newline\\\arraybackslash\hspace{0pt}}m{#1}}
\newcolumntype{C}[1]{>{\centering\let\newline\\\arraybackslash\hspace{0pt}}m{#1}}
\newcolumntype{R}[1]{>{\raggedleft\let\newline\\\arraybackslash\hspace{0pt}}m{#1}}

\newtheorem{theorem}{Theorem}[section]
\newtheorem{lemma}[theorem]{Lemma}
\newtheorem{proposition}[theorem]{Proposition}

\newtheorem{definition}[theorem]{Definition}

\newtheorem{remark}{Remark}[section]
\numberwithin{equation}{section}

\newenvironment{pfof}[1]{\vspace{1ex}\noindent{\bf Proof of #1}\hspace{0.5em}}
{\hfill\qed\vspace{1ex}}

\newcounter{arclist}

\newcounter{arcenum}
\begin{document}

	\begin{frontmatter}
		
		\title{Modelling and understanding count processes through a Markov-modulated non-homogeneous Poisson process framework}
		
		\author[UMelb]{Benjamin Avanzi}
		\ead{b.avanzi@unimelb.edu.au}

		\author[UNSW]{Greg Taylor}
		\ead{greg.taylor@unsw.edu.au}
		
		\author[UNSW]{Bernard Wong}
		\ead{bernard.wong@unsw.edu.au}
		
		\author[UNSW]{Alan Xian\corref{cor}}
		\ead{a.xian@unsw.edu.au}
		
		\cortext[cor]{Corresponding author. }
		
		\address[UMelb]{Centre for Actuarial Studies, Department of Economics \\ University of Melbourne VIC 3010, Australia}
		\address[UNSW]{School of Risk and Actuarial Studies, UNSW Business School \\ UNSW Sydney NSW 2052, Australia}
		
\begin{abstract}
The Markov-modulated Poisson process is utilised for count modelling in a variety of areas such as queueing, reliability, network and insurance claims analysis. In this paper, we extend the Markov-modulated Poisson process framework through the introduction of a flexible frequency perturbation measure. This contribution enables known information of observed event arrivals to be naturally incorporated in a tractable manner, while the hidden Markov chain captures the effect of unobservable drivers of the data. In addition to increases in accuracy and interpretability, this method supplements analysis of the latent factors. Further, this procedure naturally incorporates data features such as over-dispersion and autocorrelation. Additional insights can be generated to assist analysis, including a procedure for iterative model improvement. 

Implementation difficulties are also addressed with a focus on dealing with large data sets, where latent models are especially advantageous due the large number of observations facilitating identification of hidden factors. Namely, computational issues such as numerical underflow and high processing cost arise in this context and in this paper, we produce procedures to overcome these problems.

This modelling framework is demonstrated using a large insurance data set to illustrate theoretical, practical and computational contributions and an empirical comparison to other count models highlight the advantages of the proposed approach.
\end{abstract}
\end{frontmatter}

\section{Introduction} \label{sec:intro}
\subsection{Context}
Count modelling and the analysis of the occurrence of events is common to a wide variety of fields. The Markov-modulated Poisson process (MMPP), which is a continuous time model within the general family of Hidden Markov models (HMMs), is used in a broad spectrum of count modelling applications. It provides a natural extension upon the widely-used Poisson process in situations where an observable event process responds to external environmental stimuli. The MMPP model is composed of two components: (i) a stochastic process representing observable events over time, and (ii) a latent component that modulates the observed process. One benefit of this approach compared to the classical Poisson model stems from the greater modelling flexibility provided by the modulation, in particular to achieve over-dispersion. For example, in the context of modelling economic demand such as in \citet{NaMa15} and \citet{Art17}, the hidden Markov process proxies the external environmental drivers of demand observations which can consist of various financial, economic, social and political factors. Similarly, in insurance literature such as \citet{GuLoSt13}, claim arrivals are influenced by a variety of different environmental drivers depending on the line of business under consideration. The MMPP model provides a more faithful representation of real-world circumstances in this case because such drivers may be unobservable or extremely difficult to model, resulting in ``hidden'' states.

The MMPP model is  used in a number of other fields for comparable reasons. It is commonly found in the natural science literature to model phenomena from birth-related movements \citep{LePu92} and photon arrivals \citep{KoSuLi05} to rainfall \citep{ThRa13a} and populations \citep{LaBoSk13}. Aside from the various economic and financial areas previously mentioned, MMPPs are popular in the areas of network theory, telecommunications and data traffic modelling such as in \citet{YoKaTa01}, \citet{SaVaPa03}, \citet{ScSm03} and \citet{CaSaCr16}. Finally, this approach is utilised in the literature on risk, inventory, reliability and queueing theory to address unrealistic Poisson process assumptions through modulation, for example, in \citet{OzPa99}, \citet{Chi97}, \citet{LaOzSo13} and \citet{ArBaVa16}.

In the above papers, it is sometimes possible to apply techniques (usually Bayesian in nature) to gain inference on the latent Markov process within the MMPP. This is generally an endeavour that produces significant value, as it enhances the tractability and interpretability of the model outputs. For example, in the case of software reliability analysis as presented in \citet{LaOzSo13}, the hidden process is inferred to represent the state of the software, with labels assigned of ``excellent'', ``good'' or ``bad''. An intuitive extension upon this concept is the idea that there may exist seasonality in the modulating process which has led to the development of seasonal MMPPs in the literature, such as the model presented in \citet{GuLoSt15} for insurance claim counts where the latent process was assumed to represent the seasonal influence of various geological and climatological phenomena.

However, in any modelling exercise, there usually exist known and modellable causes of event arrivals, and these do not necessarily need to be cyclical or seasonal effects. This situation has garnered little attention in the MMPP-related academic literature, although the idea itself is not new with some discussion in the insurance context in \citet{Nor93}. Consider the case of insurance claims where the number of insurance policies under cover is clearly related to the observed claim arrivals but is unable to be taken into account as this factor is not necessarily periodic. An argument may also be made that such an intuitive driver of claims should be modelled explicitly rather than delegated to approximation through hidden regimes. Alternative examples can be easily conjured for the other fields, such as promotions producing spikes in online traffic as well as the demand (and consequently, purchases) for goods and services. Unfortunately, it is difficult to structure this information into the standard MMPP model, as regime selection is generally an automatic procedure. 

\subsection{Contributions}
In this paper, an extension to the standard MMPP model is introduced which includes a component to explicitly control for the frequency perturbation caused by \emph{known} factors. This measure is chosen to be very flexible, so that \emph{both} cyclical and non-recurring trends can be captured. The introduced heterogeneity here produces the Markov-modulated \emph{non-homogeneous} Poisson process (MMNPP).

The advantages of the proposed framework are many. Firstly, the original benefits of MMPPs in analysing count processes that are outlined above are retained. Secondly, the distinct separation of the known and hidden drivers of event arrivals enhances the insights obtained. This greatly benefits the utility of the approach, as inappropriate or insufficient consideration of the known components can confound any inferences obtained about the unobservable regimes of the hidden Markov chain. By explicitly taking these factors into account, the modelling outputs may become more interpretable and this is demonstrated in an empirical case study in Section \ref{sec:CaseStudy} where the MMNPP model is utilised as a diagnostic tool for generating inferences which further assist the modelling procedure. Data features such as over-dispersion, auto-correlation and persistence in regimes are also able to taken into account in an understandable manner. This included information, in addition to the domain knowledge structured into the model, can ultimately lead to gains in accuracy and precision.

While non-homogeneous extensions to the MMPP model have not been widely explored in the literature (see \citet{ThRa13b,GuLoSt15} for the examples that the authors are aware of), comparable models have been successfully utilised for the more general family of Hidden Markov models. In these cases, the latent model is implemented to capture various unobservable factors with the non-homogeneous component supplementing this analysis. For example, \citet{MoNeJe10} apply a non-homogeneous HMM to assess the effects of pharmaceutical marketing activities on physicians, taking into account their heterogeneity and behaviour dynamics. \citet{ShZh14} apply a non-homogeneous HMM to investigate decision aids for online purchases with time-varying parameters across the latent Markov states. Further examples of applications include buyer-seller dynamics in \citet{ZhNeAn14} as well as the modelling of phenomena such as precipitation \citep{HuGuCh99, GrRoSmTr11} and pollution concentration \citep{LaMaPi11}. This approach is also found in insurance analysis in \citet{BaChLiTa19}, who use a Pascal-HMM to model claim arrivals. Analogously, our proposed non-homogeneous extension will provide a large amount of flexibility and practicability to the MMNPP model in order to improve analysis in the varied areas where it is applied. Further, in comparison to the HMM approach, our model does not require considerations of discretisation as both the arrival processes and the underlying Markov chain are continuous. It is noted in some papers such as \citet{BaChLiTa19} and \citet{CrAnVe19} that choice of granularity and appropriate assumptions can be a non-trivial task, and the time-scale invariance of the continuous approach is advantageous.

Another contribution in this paper is that we show that the MMNPP is a specific operational time transform of the standard MMPP. The form of the transform depends on the frequency perturbation measure which provides much intuition to the modelling process which aids the interpretability of the final model outputs.

Finally, implementations found in the literature generally utilise examples with sample sizes that are relatively small. While this may be appropriate in certain areas, consideration of larger data sets is of merit, particularly given the current popularity of analytic models that are applied to such data sets. Further, the large number of observations advantages latent models such as the MMNPP due to clearer identification of hidden regimes. 

Under these circumstances, calibration problems such as long computation times and numerical instability are greatly exacerbated. Without addressing these problems, applicability of the proposed approach is more limited. However, it is exactly these situations that are ideal for analysis using hidden Markov models as the unobservable regimes are more clearly identified. In this paper, we produce an Expectation-Maximisation (EM) calibration algorithm that introduces several computational developments to reduce calibration time, allowing the model to be calibrated in a reasonable period of time when there is a large amount of data to be processed. Further, the issue of numerical underflow is dealt with through a scaling procedure applied to the forward-backwards recursions. These practical considerations and contributions are outlined in the next section with context provided to demonstrate the necessity of these developments, as otherwise, the model is unable to be implemented and thus provides little utility.

In the following section, we outline the insurance claim analysis context that serves as a motivating example for the developments within this paper. The MMNPP model specifications are then presented in Section \ref{sec:ModelSpec}. In Section \ref{sec:EMCal}, we develop a calibration procedure including various techniques to alleviate the computational problems explained previously. In Section \ref{sec:CaseStudy}, we implement the model on a real data set to highlight the inferential capabilities of our proposed approach. We also provide a comparison to other models to highlight the advantages provided by the MMNPP. Section \ref{sec:Conc} concludes.

\subsection{Detailed motivating example: Insurance claim counts} \label{S_ins}
The empirical case study in Section \ref{sec:CaseStudy} examines daily claim counts for a motor insurance line of business for a large Australian non-life insurer. The analysis of insurance claim occurrences is a common procedure and serves to inform many insurance business operations. There is a large amount of actuarial science literature on this topic but until recently, analysis has exploited data aggregated over longer discrete time frames such as months or years. Weekly and daily granularity of claim count data has been investigated by only a few papers such as \citet{VeWu16, AvWoYa16, CrAnVe19} which deal with the reporting delay aspect of insurance claims while other nuances such as seasonality are demonstrated in \citet{GuLoSt15, AlArBe19}. The benefits of analysis at this greater level of detail include the generation of insights that were previously masked by aggregation and discretisation, which can lead to greater accuracy and precision.

Insurance claim count modelling encounters several obstacles that are significantly alleviated by the innovations made in this paper. Particularly for well known lines of business such as Motor, insurers possess established domain knowledge of important covariates that influence claim observations and the inability to incorporate this information into analysis would hamper to utility of any such modelling approach. Secondly, analysis at greater granularity results in a very large number of individual data points as opposed to the original aggregated observations (significantly more than half a million in our real data case study). This aggravates calibration issues common to hidden Markov models such as numerical instability as well as considerations such as calibration time and processing power required. Our developments allow the model to be calibrated on standard computing software in a reasonable amount of time. Finally, the insights  in Section \ref{sec:CaseStudy} highlight the diagnostic abilities of the modelling framework, including the capacity for iterative model improvement. This capability is particularly relevant for insurance claim modelling as analysis at this level of detail is not commonly undertaken and we demonstrate that additional interpretable causes of claim observations are able to be extracted. These drivers can then be examined and incorporated as deemed fit by the modeller.

\section{Model Specifications for the MMNPP} \label{sec:ModelSpec}
We assume that event arrivals, over the time interval $[0,T]$, follow a Markov-modulated non-homogeneous Poisson process $N_M = \{N_M (t); t\ge 0\}$. Here, $N_M(t)$ represents the total number of events arriving before time $t$, with $N_M(0)=0$.  There is also assumed to be an unobservable underlying environmental process that impacts the event arrival intensity. This process is modelled with a continuous time Markov chain $M(t)$ with finite discrete state space ${1, 2, \ldots, r}$. We denote this latent process $M = \{M(t);t\ge 0\}$, and it is assumed that the chain has constant stationary transition intensities between states $i$ and $j$ of $q_{i,j}$. Consequently, if the $n$-th state change time is denoted by $u_n$, then the duration of the $n$-th state is exponentially distributed and given by
\begin{equation}
\mathbb{P}[u_{n+1} - u_{n} > t |  M(u_{n})=i ] = \exp \left(- q_{i} t \right), \text{ where } q_i =  \sum_{j \ne i} q_{i,j},
\end{equation}
where $\mathbb{P}$ is the probability function.

When the Markov process is in state $i$ at time $t$, events arrive according to the process $N_M$ with intensity rate 
\begin{equation}
\lambda_{M}(t) = \lambda_{M(t)} \times \gamma(t).
\end{equation}
Here, $\lambda_{M(t)}$ is the constant intensity conditional on the state of the underlying Markov chain at time $t$, and $\gamma$ represents the known exogenous volume or exposure process. The latter is at the heart of the contributions of this paper, as discussed in detail in the next section. This $\gamma$-process is very flexible and can take the form of any desired function of bounded variation which covers the great majority of all possible functions that may arise in practical modelling circumstances. 

The standard expressions for the conditional Poisson processes are easily obtained under this framework. The probability of $n$ events occurring before time $t$ is given by
\begin{equation}
\mathbb{P}[N_M (t) = n | M(s); 0 \le s \le t] = \exp \left(-\int_{0}^{t} \lambda_{M(s)} \gamma(s) ds \right) \frac{\left(\int_{0}^{t} \lambda_{M(s)} \gamma(s) ds\right)^n}{n!}.
\end{equation}

From this expression, properties of the inter-arrival times of events may also be obtained. Let $t_n$ be the arrival time of the $n$-th event. Then 
\begin{equation}
\mathbb{P}[t_{n+1} > t_n + t | t_{n},  M(s); t_n \le s \le t_n + t ]= \exp \left(-\int_{t_{n}}^{t_{n}+t} \lambda_{M(s)} \gamma(s) ds \right).
\end{equation}
In the paper, the overall MMNPP process to describe the hidden regimes and event arrivals is denoted as $\{M,N_M\} = \{M(t),N_M(t); t \ge 0\}$.

\subsection{The exposure measure $\gamma$}
In the above, $\gamma(t)$  is a scalar function representing the known exogenous volume or exposure process.  This $\gamma$-process is very flexible and theoretically only needs to be a function of bounded variation, which allows for both jump-type and removable discontinuities (although the latter would rarely occur in practice). For the purposes of calibration using the method outlined in Section \ref{sec:EMCal}, we require that any applicable function can be adequately approximated by a number of constant-valued intervals. This is a fairly weak assumption as the simple function approximation is very general, allowing for a wide range of functions to be used such as continuous and finitely dis-continuous functions. It would be difficult to envisage a volume process in any practical circumstance that would be suitably irregular to violate this assumption. One example of such an extreme case would be a function consisting of purely discontinuous singletons on the real line, but we could not think of any practical situation where the volume function would not satisfy our requirements. The reasoning behind this constraint is that it greatly reduces the computational burden in evaluating several complex expressions involving matrix differential equations. This is detailed within the proof for Theorem \ref{thm:CKthm1} in Section \ref{sec:Estep}. The multiplicative structure of the overall Poisson intensity is also theoretically justified in Section \ref{sec:ExpJust} but it is intuitively explainable. For example, if the number of consumers within the market for a certain good doubles, then it is reasonable to expect that the observed conditional Poisson intensity of purchases of the good also doubles, ceteris paribus.

\subsection{An operational time scaling interpretation}
The MMNPP can be interpreted as a operational time transformation of the a homogeneous MMPP, which provides a convenient, intuitive interpretation of the model. This argument follows in a similar manner to the transformation applicable to non-homogeneous Poisson processes and is outlined below.

We begin with the definition of an operational time scale that will be applied.
\begin{definition} \label{def:optimedef}
	Consider an operational time scaling function $\rho$ applied to time $t$, where $\rho(t)$ is a monotonically increasing function in $t$ and $\rho(0) = 0$. As a result, $\rho^{-1}(t)$ is also well defined. This produces the operational time scale $\rho(t)$. Using $\rho$, we can define a new time-transformed process $N_M(\rho^{-1}(t))$:	

	\begin{align}
		t 		& \rightarrow \rho^{-1}(t) \nonumber \\
		N_M(t) 	& \rightarrow N_M(\rho^{-1}(t)) \nonumber
	\end{align}
\end{definition}

We will also use the following specific operational time function $\rho$:

\begin{definition}
	In the context of the MMNPP model, the operational time function $\rho$ takes the form
	\begin{equation*} \label{eq:optimetrans}
	\rho(t) = \int_{0}^{t} \gamma(x) dx.
	\end{equation*}
\end{definition}

It is shown in Theorem \ref{thm:OpTimeTransHomo} below that the original MMNPP process $\{M(t),N_M(t); t \ge 0\}$ under this transformation becomes $\{M(t), N_M(\rho^{-1}(t)) ; t \ge 0 \}$, where the component $N_M(\rho^{-1}(t))$ is a homogeneous Poisson process with constant conditional intensity rate $\lambda_{M(t)}$ at time $t$. Observe that in the above, the time scaling has only been applied to the conditional Poisson process and not the hidden Markovian process.

\begin{theorem}\label{thm:OpTimeTransHomo}
	We use the previously defined time transform from $t$ to $\rho^{-1}(t)$:
	\[
	\underset{\text{old time}}{t} \rightarrow  \underset{\text{new time}}{\rho^{-1}(t)},
	\]
	where $\rho(t) = \int_{0}^{t} \gamma(x) dx$ is a monotonically increasing function in $t$ with $\rho(0) = 0$. This transform is applied to the conditional Poisson component of a MMNPP process $\{M(t),N_M(t);0\le t \le T\}$ with transitions $q_{i,j}$ and intensities $\lambda_{M(t)}\times \gamma(t)$. The resultant process is the homogeneous MMPP process $\{M(t),N_M(\rho^{-1}(t));0\le t \le T\}$ with transitions $q_{i,j}$ and intensities $\lambda_{M(t)}$.
\end{theorem}

\begin{pfof}{Theorem \ref{thm:OpTimeTransHomo}}
	Although the result is very intuitive, for completeness we provide a proof here. We require the following preliminary results. For readability, the proofs of the Lemmas are provided instead in the appendix.
	
	\begin{lemma}\label{lem:MMNPPmarkov}
		The time transformed process $\{M(t),N_M(\rho^{-1}(t))\}$ retains the Markov property
	\end{lemma}
	\begin{pfof}{Lemma \ref{lem:MMNPPmarkov}}
		Please refer to Appendix \ref{app:lem1}.
	\end{pfof}

	\begin{lemma}\label{lem:MMNPPiffcond}
		The MMPP process $\{ M(t),N_M(t) \}$ is homogeneous over period $[t_1,t_2]$ if and only if
		\begin{equation*}
		\mathbb{P} \left[ N_M(t_1) = n, N_M(t_2) =n \right] = \exp \left[ -\sum_{k=1}^{m} \lambda_{s_k} (u_{k} - u_{k-1}) \right]
		\end{equation*}
		where $k$ is a counter for the number of regime periods that occur in $[t_1,t_2]$, $s_k$ is the ordinal number of the $k$-th regime and regime changes occur within this interval at times $u_1, u_2, \ldots, u_m$ with $u_0 = t_1$ and $u_m = t_2$.
	\end{lemma}
	\begin{pfof}{Lemma \ref{lem:MMNPPiffcond}}
		Please refer to Appendix \ref{app:lem2}.
	\end{pfof}

	In order to satisfy the non-decreasing property of $\rho$, $\gamma$ is restricted to being strictly positive (which is an intuitively correct restriction for an exposure measure). Using $k$ as the number of regime changes between $[t_1,t_2]$, we obtain
	\begin{align}
	\mathbb{P} \left[ N_M(\rho^{-1}(t_1)) = n, N_M(\rho^{-1}(t_2)) = n \right] &= \mathbb{P} \left[ N_M(t_1) = n, N_M(t_2) = n \right] \\
	&= \exp \left[ -\int_{t_1}^{t_2} \lambda_{M(x)} \times \gamma(x) dx \right] \\
	&= \exp \left[ - \sum_{k=1}^{m} \int_{u_{k-1}}^{u_{k}} \lambda_{s_k} \times \gamma(x)  dx \right] \\ 
	&= \exp \left[ -\sum_{k=1}^{m} \lambda_{s_i} \int_{u_{k-1}}^{u_k} \gamma(x) dx \right] \\
	&= \exp \left[ -\sum_{k=1}^{m} \lambda_{s_k} [\rho(u_k) - \rho(u_{k-1})] \right]
	\end{align}
	
	The proof is concluded with the observation that the times $\rho(u_k)$ correspond to the regime change times $u_k$ in the underlying Markov chain $M$. Thus, using Lemmas \ref{lem:MMNPPmarkov} and \ref{lem:MMNPPiffcond}, the operational time adjustment as defined above changes the original non-homogeneous MMPP into a homogeneous MMPP when the non-homogeneous intensities are in the form $\lambda(t) = \lambda_{M(t)} \times \gamma(t)$.
\end{pfof}

\subsection{Intuition for the form of the exposure measure}\label{sec:ExpJust}
We have defined the non-homogeneous Poisson intensity $\lambda_{M}(t) = \lambda_{M(t)} \times \gamma(t)$, where $\gamma(t)$ was a known exogenous process that reflected external risk exposures. In addition to the explanation in the previous section, The form of this intensity can be theoretically and intuitively justified in the following manner. If we have multiple (homogeneous) MMPP processes with the same modulating Markov process, then we can show that the standard additivity property of Poisson processes still holds. This result follows directly from Proposition \ref{prop:addMMNPP} below.
\begin{proposition} \label{prop:addMMNPP}
	Let there be two MMPP processes $\{M(t),N^1_{M} (t)\}$ and $\{M(t),N^2_M (t)\}$ with the same underlying Markov chain $\{M(t)\}$ and claim intensities $\lambda^1_{M}(t)$ and $\lambda^2_{M}(t)$ respectively. We also have that the conditional Poisson processes $N^i_{M} (t)$ are independent for $i = 1,2$. Then the combined process
	\[
	\{M(t),N^{1,2}_M (t)\} = \{M(t),N^1_M (t) + N^2_M (t)\}
	\]
	is also an MMPP with a conditional Poisson process $N_M^{1,2} (t)$ that has a claim intensity of $\lambda_{M}(t) = \lambda^1_{M}(t) + \lambda^2_{M}(t)$.
\end{proposition}

\begin{pfof}{Proposition \ref{prop:addMMNPP}}
	Please refer to Appendix \ref{app:prop1}
\end{pfof}

Generalising Proposition \ref{prop:addMMNPP} to $n$ event processes of form $\{ M(t), N_M^i(t) \}_{i=1,\ldots,n}$ with the same intensity $\lambda_{M(t)}$ (for example, risk events arriving from $n$ similar insurance policies or arrivals to $n$ similar queues at airport check-in counters), the combined MMPP process has a claim intensity of $\lambda_{M(t)} \times n$. If this exposure process is allowed to vary over time, then the event intensity function can be thought of as $\lambda_{M(t)} \times \gamma(t)$, where $\gamma(t)$ is the volume or risk exposure measure. This interpretation also applies when the known function $\gamma$ is a combination of multiple factors, in which case the multiplicative assumption is convenient. In this case, the overall event intensity would also be of the form $\lambda_{M(t)} \times \gamma(t)$, where $\gamma$ is now the multiplicative combination of the impact of the considered factors.

\section{Calibration of the MMNPP framework}\label{sec:EMCal}
The calibration of the MMNPP model is a two-step procedure and is detailed in the following section. The steps involved are

\begin{enumerate}
	\item Calibration of the exposure/volume component $\gamma(t)$.
	\item Calibration of the hidden Markov-modulated Poisson component, specifically event frequency and regime transition parameters $\lambda_M$ and $q_{i,j}$ respectively.
\end{enumerate}

\subsection{Calibration of the exposure/volume component $\gamma$}
Firstly, modellers will combine their domain knowledge and data analysis to generate a model for the exposure/volume component of the framework, which represents the known or ``explicitly modellable" information. One key advantage of our proposed approach is that this component is modular and many different models can be chosen depending on the modelling context. For example, for the purpose of modelling insurance claims, generalised linear models (GLMs) are commonly applied in practice, although there has been recent exploration into the usage of more sophisticated data-driven methods such as trees and neural networks such as in \citet{GaRiWu19}. All such methods are applicable here due to the flexibility of the measure $\gamma$. Note that without loss of generality, the scalar process $\rho$ is defined relative to some base level, such as the exposure/volume at the beginning of the period.

The values obtained from the exposure component of the framework then serve as inputs in the hidden Markov chain calibration. Intuitively, the hidden Markov chain captures any residual temporal effects that were not explicitly included in the model for the exposure component. The details for this secondary step in the calibration is provided in the following section.

\subsection{Calibration of the hidden Markov component using the EM algorithm}

The calibration of the hidden Markov component of the MMNPP model utilises an adapted form of the Expectation Maximization algorithm for MMPPs originally proposed in \citet{Ryd94}. Alternative methods exist in the literature for the homogeneous MMPP models such as moment-matching \citep{Ros89, Gus91}, other likelihood-based estimation approaches \citep{Mei84} and Markov chain Monte Carlo (MCMC) sampling, the last of which has been popular in the literature in recent times. \citet{Ryd08} examine the EM vs MCMC estimation for the general family of Hidden Markov models (HMM). The paper takes a computational perspective, and concludes that in the case where only point estimates are required or model comparison only requires likelihoods, the EM algorithm is preferred. In other cases, either approach has advantages and disadvantages and the comparison is less clear.

When applying the model to large data sets, as is the case in Section \ref{sec:CaseStudy}, the relative computational simplicity of the EM algorithm is beneficial. For MMPPs, closed form expressions for the estimators of the algorithm can be obtained, alleviating the high computational requirements. We extend those results to the MMNPP case in this paper. Additional improvements to the algorithm have been made to further reduce computation times and to resolve the previously mentioned issue of numerical instability. 

A very brief summary of one loop of the iterative algorithm is the following (note that the estimators and expressions for their computation will be introduced in detail later):

\begin{enumerate}
	\item Using current estimates of the regime transition and regime frequency parameters  $Q$ and $\Lambda$ respectively, evaluate the E-step estimators $\boldsymbol{\hat{a}}, \boldsymbol{\hat{n}}, \boldsymbol{\hat{T}}$ and $\boldsymbol{\hat{T}^*}$.
	\item Using current estimates of the E-step estimators $\boldsymbol{\hat{a}}, \boldsymbol{\hat{n}}, \boldsymbol{\hat{T}}$ and $\boldsymbol{\hat{T}^*}$, evaluate new estimates for $Q$ and $\Lambda$.
	\item Stop the algorithm based on some termination criteria (more details in Section \ref{sec:EMsum}).
\end{enumerate}

\vspace{0.5em}

The following sections are organised as follows. In Section \ref{sec:M-step}, the complete likelihood function that is maximised through the EM algorithm is derived, taking into account the introduced non-homogeneity. In Section \ref{sec:Estep}, the closed form expressions for the E-step estimators given with derivations provided in the appendix. In order to overcome numerical instability issues, a scaling approach is applied to the recursion equations, as outlined in Section \ref{sec:ScaledRec}. This approach is adapted from the algorithm for MMPPs from \citet{RoEpDi06}. We also adapt a result from \citet{Van78} to increase computational efficiency when evaluating integrals of matrix exponentials, which was previously the most computationally burdensome component of the algorithm. If the algorithm was instead forced to rely on quadrature methods for these evaluations, for example, the calculations would quickly become too cumbersome for even moderate numbers of claims to be viably analysed. Finally, during the maximisation step shown in Section \ref{sec:M-step}, the traditional time-based estimators presented in \citet{Ryd96} are now split into a time-based and operational time-based estimator, which follows intuitively from the operational time transform described in the previous section. Section \ref{sec:EMsum} concludes with a brief summary of the iterative calibration procedure.

\subsection{Complete likelihood function of an MMNPP}\label{sec:M-step}
The EM algorithm is an iterative procedure that seeks to maximise the complete likelihood function (that is, the likelihood function assuming knowledge of the hidden components). In the following section, we produce the complete likelihood function as well the necessary E-step estimators for the algorithm. 

The complete form of likelihood function in the case of the MMNPP model, assuming knowledge of both event arrival times and regime change times, is what needs to be maximised in this case. We introduce the following notation:
\begin{enumerate}
	\item Jumps between states/regimes occurs at time points $0=u_0<u_1<u_2<\ldots<u_m<u_{m+1}=T$, where $T$ is the end of the observation period.
	\item The regime time interval is $I_k = [u_{k-1},u_k)$ for $1 \le k \le m+1$.
	\item The regime of the process $M(t)$ in the interval $I_k$ is given by $s_k \in \{1,2,\ldots, r\}$.
	\item The number of events occurring in regime period $I_k$ is given by $z_k \in \mathbb{N}$.
	\item The total number of events is denoted $n$.
	\item Event arrivals occur within $I_k$ at time points $t_{k,1}, t_{k,2}, \ldots, t_{k,z_k}$ and $t_{k,0} = u_{k-1}$. 
\end{enumerate}

Further, the following result is required in order to derive the complete likelihood function:
\begin{lemma}\label{lem:distclaims}
	Consider the regime time interval $I_k = [u_{k-1},u_k)$ of a non-homogeneous MMPP with claim intensity $\lambda(t) = \lambda_{s_k} \times \gamma(t)$ where the underlying Markov chain $M(t)$ does not change states, and $\lambda_{s_k}$ is the constant component of the claim intensity within $I_{k}$. The joint density of claim times $T_i$, given that $z_k$ claims occur within this interval, is given by the following expression
	\begin{equation*}
	\mathbb{P}\left[ T_{k,1} = t_{k,1}, T_{k,2} = t_{k,2}, \ldots, T_{k,z_k} = t_{k,z_k} | N_M(u_k) - N_M(u_{k-1}) = z_k \right] = \frac{z_k!}{\left(\int_{u_{k-1}}^{u_{k}} \gamma(s) ds \right)^{z_{k}}} \prod_{i=1}^{z_k} \gamma(t_{k,i})
	\end{equation*}
\end{lemma}

\begin{pfof}{Lemma \ref{lem:distclaims}}
It can be seen that
	\begin{align*}
	& \mathbb{P}\left[ T_{k,1} = t_{k,1}, T_{k,2} = t_{k,2}, \ldots, T_{k,z_k} = t_{k,z_k} | N_M(u_k) - N_M(u_{k-1}) = z_k \right] \\
	& =  \frac{\mathbb{P}\left[ T_{k,1} = t_{k,1}, T_{k,2} = t_{k,2}, \ldots, T_{k,z_k} = t_{k,z_k}, N_M(u_k) - N_M(u_{k-1}) = z_k \right]}{\mathbb{P}[N_M(u_k) - N_M(u_{k-1})=z_k]}
	\end{align*}
	Note that the numerator of the above can be rewritten in the following manner:
	\begin{align*}
	& \mathbb{P}\left[ T_{k,1} = t_{k,1}, T_{k,2} = t_{k,2}, \ldots, T_{k,z_k} = t_{k,z_k} , N_M(u_k) - N_M(u_{k-1})=z_k \right] \\
	& = \mathbb{P}\left[z_k \text{ claims between } \left[u_{k-1},u_k \right) \text{ with claim times } t_{k,1},t_{k,2},\ldots,t_{k,z_k}\right] \\
	& = \prod_{i=1}^{z_k} \bigg[ \mathbb{P}\left[ \text{ no claims until time } t_{k,i} \right] \times \mathbb{P} \left[ \text{ claim at time } t_{k,i} \right] \bigg] \times \mathbb{P}\left[ \text{ no claims until end of interval } u_k \right]\\
	& = \prod_{i=1}^{z_k} \left[ \exp \left(-\int_{t_{k,i-1}}^{t_{k,i}} \lambda_{s_k} \gamma(y) dy \times \lambda_{s_k} \right) \gamma(t_{k,i}) \right] \times \exp \left( -\int^{u_k}_{t_{k,z_k}} \lambda_{s_k} \gamma(y) dy \right)
	\end{align*}
	Thus, we have that
	\begin{align*}
	& \mathbb{P}\left[ T_{k,1} = t_{k,1}, T_{k,2} = t_{k,2}, \ldots, T_{k,z_k} = t_{k,z_k} | N_M(u_k) - N_M(u_{k-1}) = z_k \right] \\
	& =  \frac{\prod_{i=1}^{z_k} \left[ \exp \left(-\int_{t_{k,i-1}}^{t_{k,i}} \lambda_{s_k} \gamma(y) dy \right) \lambda_{s_k}  \gamma(t_{k,i}) \right] \times \exp \left( -\int^{u_k}_{t_{k,z_k}} \lambda_{s_k} \gamma(y) dy \right)}{\exp\left( - \int^{u_k}_{u_{k-1}} \lambda_{s_k} \gamma(y) dy \right) \frac{(\int^{u_k}_{u_{k-1}} \lambda_{s_k} \gamma(y) dy )^{z_{k}}}{z_{k}!}} \\ 
	& = \frac{z_k!}{\left(\int_{u_{k-1}}^{u_{k}} \gamma(s) ds \right)^{z_{k}}} \prod_{i=1}^{z_k} \gamma(t_{k,i}).
	\end{align*}
\end{pfof}

The complete likelihood function of the MMNPP is then given in Equation \eqref{eq:CompleteLikelihood}, using Lemma \ref{lem:distclaims} to determine the last term:
\begin{align}
\mathcal{L}^{c} & = \pi_{s_{0}}\left\{ \prod_{k=1}^{m}q_{s_{k}}\exp\left(-q_{s_{k}}\Delta u_{k}\right)\times\frac{q_{s_{k},s_{k+1}}}{q_{s_{k}}}\right\} \exp\left(-q_{s_{m+1}}\Delta u_{m+1}\right)\times \nonumber \\
&  \left\{ \prod_{k=1}^{m+1}\frac{\left(\int_{u_{k-1}}^{u_k} \lambda_{s_{k}} \gamma(t) dt \right)^{z_{k}}}{z_{k}!}\exp\left(-\int_{u_{k-1}}^{u_k} \lambda_{s_{k}} \gamma(t) dt \right) \times \frac{z_k!}{\left(\int_{u_{k-1}}^{u_{k}} \gamma(t) dt \right)^{z_{k}}} \times \prod_{i=1}^{z_k} \gamma(t_{k,i}) \right\} \\
& = \pi_{s_{0}}\left\{ \prod_{k=1}^{m} \exp\left(-q_{s_{k}}\Delta u_{k}\right)\times q_{s_{k},s_{k+1}} \right\} \exp\left(-q_{s_{m+1}}\Delta u_{m+1}\right)\times \nonumber \\
& \left\{ \prod_{k=1}^{m+1} \left[ \lambda^{z_k}_{s_k} \exp\left(-\int_{u_{k-1}}^{u_k} \lambda_{s_{k}} \gamma(t) dt \right) \prod_{i=1}^{z_k} \gamma(t_{k,i}) \right] \right\} \label{eq:CompleteLikelihood}
\end{align}

In the above, $\pi_{s_{0}}$ denotes the starting probabilities of the underlying Markov chain. If prior information is unavailable, a discrete uniform distribution is applicable. By taking logs, we obtain
\begin{align} \label{eq:CompleteLoglikelihood}
\log \mathcal{L}^{c} & = \log	\pi_{s_1} - \sum_{k=1}^{m+1} q_{s_{k}} \Delta u_k + \sum_{k=1}^{m} \log q_{s_k , s_{k+1}} \nonumber \\
& \quad + \sum_{k=1}^{m+1} z_k \log \lambda_{s_k} - \sum_{k=1}^{m+1} \lambda_{s_{k}} \int_{u_{k-1}}^{u_k} \gamma(t) dt + \sum_{k=1}^{m+1} \sum_{i=1}^{z_k} \log(\gamma(t_{k,i})). 
\end{align}

These sums are then reformulated in terms of the order of the Markov chain $r$ instead of the number of regime changes. Note that the last term in Equation \eqref{eq:CompleteLoglikelihood} is the summation of all the logs of the exposures at the claim arrival times. As these values are known, this term is rewritten as the constant $C(\gamma)$. 

We also require the following E-step estimators, which are generally analogous to the estimators from \citet{Ryd94}. The first two, $m_{i,j}$ and $n_{i}$ can be interpreted to the number of transitions from state $i$ to state $j$ and the number of claims arriving in state $i$ respectively.
\begin{align}
a_{i,j} & = \text{card}\{ t: 0 < t \le T, N_M(t-) = i, N_M(t) = j \} \\
n_{i} 	& = \sum_{k=1}^{n} \mathbb{I}_{\{M(t_k) = i\}},
\end{align}
where $\text{card}$ is a count/cardinality function for the number of events. The total amount of time spent in state $i$ is
\begin{equation}
T_i = \int_{0}^{T} \mathbb{I}_{\{M(t)=i\}}dt.
\end{equation}
Compared to the original EM algorithm, there is an extra ``time in state'' estimator $T^*_i$, given by 
\begin{equation}
T^*_i = \int_{0}^{T} \mathbb{I}_{\{M(t)=i\}} \gamma(t) dt,
\end{equation}
which reduces to the definition of $T_i$ in the case of a homogeneous MMPP where $\gamma(t) = 1 \, \forall t$. Finally, the complete loglikelihood expression may be written as
\begin{equation}\label{eq:finalcomploglik}
\log \mathcal{L}^{c} = \sum_{i=1}^{r} \mathbb{I}_{\{M(0)=i\}} \log \pi_i - \sum_{i=1}^{r} T_i q_i + \sum_{i=1}^{r} \sum_{j=1, i \ne j}^{r} a_{i,j} \log q_{i,j} + \sum_{i=1}^{r}  n_i \log \lambda_i - \sum_{i=1}^{r} \lambda_i T^*_i + C(\gamma)
\end{equation}
It is interesting to note that the forms of $T_i$ and $T^*_i$ point to the use of two different time scales to evaluate the MMNPP parameters $q$ and $\lambda$, where the relationship between the different time scales is defined through the exposure measure $\gamma$. This is intuitive given our previous operational time interpretation of the MMNPP process.

\subsection{E-step of the adapted EM algorithm for MMNPPs}\label{sec:Estep}
In the next section, we will derive closed form expression for the expectation of the estimators defined above. We require the evaluation of 
\begin{equation}
\mathbb{E}_{(Q^0,\Lambda^0)}[\log \mathcal{L}^c (Q,\Lambda)| N_M(\rho^{-1}(t)), 0 \le t \le T].
\end{equation}
In the above, $Q = \{q_{i,j}\}$, $\Lambda = \text{diag}(\lambda_{M_1},\lambda_{M_2},\ldots,\lambda_{M_r})$ and the current estimates of the parameters of the MMNPP at each iteration of the EM algorithm are denoted by $q^0_{i,j}$ and $\lambda^0_{i}$. Their respective matrices are correspondingly $Q^0$ and $\Lambda^0$. For notational convenience, we will also rewrite the $t_{k,i}$ terms in chronological order as $t_{j},\, j=1,\ldots,n$. The following expressions for the expectations of the M-step estimators are obtained:
\begin{alignat}{3}
&\hat{a}_{i,j} 	&& = \mathbb{E}[a_{i,j}|N_M(\rho^{-1}(t)), 0 \le t \le T] && = \int_{0}^{T} \mathbb{P} [M(x-)=i,M(x)=j | N_M(\rho^{-1}(t)), 0 \le t \le T] dx \label{eq:miestimator} \\
&\hat{n}_{i}		&& = \mathbb{E}[n_{i}|N_M(\rho^{-1}(t)), 0 \le t \le T] && = \sum_{k=1}^{n} \mathbb{P} [M(t_{k}) = i | N_M(\rho^{-1}(t)), 0 \le t \le T] \label{eq:niestimator}\\
&\hat{T}_i		&& = \mathbb{E}[T_{i}|N_M(\rho^{-1}(t)), 0 \le t \le T] && = \int_{0}^{T} \mathbb{P} [M(x) = i | N_M(\rho^{-1}(t)), 0 \le t \le T ] dx \label{eq:tiestimator} \\
&\hat{T}^*_i		&& = \mathbb{E}[T^*_{i}|N_M(\rho^{-1}(t)), 0 \le t \le T] && = \int_{0}^{T} \mathbb{P} [M(x) = i | N_M(\rho^{-1}(t)), 0 \le t \le T ] \gamma(x) dx \label{eq:tistarestimator}
\end{alignat}

The above expressions are decomposed into event inter-arrival intervals and written in terms of the fundamental regime transition probabilities for MMNPPs within each interval. These probabilities are be derived from the Chapman-Kolmogorov equations, taking into account the varying exposure measure at event arrival times, and are given in Theorems \ref{thm:CKthm1} and \ref{thm:CKthm2} below.

\begin{theorem}\label{thm:CKthm1}
	The probability of regime transitions without any event arrivals by time $t+t^*$, given the regime at time $t$, is given (in matrix form) by
	\begin{equation}
	\boldsymbol{\bar{F}}(t,t^*) = \exp \left[ \left(Q-\Lambda \gamma_{t^*} \right) t^* \right],
	\end{equation}
	where $\gamma_{t^*}$ is the constant exposure measure that applies to the interval $[t,t+t^*)$.
\end{theorem}

\begin{pfof}{Theorem \ref{thm:CKthm1}}
The transition probabilities without claim arrivals at time $t^*+t$ given the state at time $t$ is given by 
\begin{align*}
\bar{F}_{i,j} (t,t^*) & = \mathbb{P} [M(t^* + t)=j,N_M(\rho^{-1}(t^* + t))=n|M(t^*)=i,N_M(\rho^{-1}(t^*))=n]
\end{align*}
for the process $\{ M(t), N_M(\rho^{-1}(t)) \}$, given that $n$ claims have occurred before time $t^*$. For any arbitrarily small increment $\Delta t$,
\begin{equation*}
\bar{F}_{i,j}(t+\Delta t,t^*) = \bar{F}_{i,j}(t,t^*) (1-\sum_{i\ne j}^{r} q_{j,i} \Delta t - \lambda_j \Delta \rho(t^* + t) ) + \sum_{k \ne j}^{r} \bar{F}_{i,k} (t,t^*) q_{k,j} \Delta t + o(\Delta t)
\end{equation*}
where $\Delta \rho(t^* + t) = \rho(t^* + t+\Delta t) - \rho(t^* + t)$. Taking the limit as $\Delta t$ goes to zero and using the previously specified constraint that $q_i = \sum_{j\ne i} q_{i,j}$, the following Chapman-Kolmogorov differential equations are obtained:
\begin{equation}\label{eq:CK1}
\bar{F}'_{i,j}(t,t^*) = \bar{F}_{i,j} (t,t^*) (-q_{j}-\lambda_j \lim_{\Delta t \rightarrow 0} \frac{\Delta \rho(t^* + t)}{\Delta t}) + \sum_{k \ne j}^{r} \bar{F}_{i,k} (t,t^*) q_{k,j}
\end{equation}
Note that the limit in the above equation is simply $\rho'(t^* + t)$, which equals $\gamma(t^* + t)$ from the Fundamental Theorem of Calculus. Then Equation \eqref{eq:CK1} may be rewritten as
\begin{equation*}
\bar{F}'_{i,j}(t,t^*) = \bar{F}_{i,j} (t,t^*) (-q_{j}-\lambda_j \gamma(t^* + t)) + \sum_{k \ne j}^{r} \bar{F}_{i,k} (t,t^*) q_{k,j}
\end{equation*}
which in matrix form is $\boldsymbol{\bar{F}'}(t,t^*) =  \boldsymbol{\bar{F}}(t,t^*) (Q-\Lambda \gamma(t^* + t)).$

The assumption that $\gamma(t)$ is constant between event arrivals and exposure component changes is applied here so that the above matrix differential equation may be simplified. In the proposed EM algorithm, each part of $\bar{F}_{i,j}$ is evaluated over disjoint intervals where there are no event arrivals or changes in the exposure component. In this circumstance, $\gamma(t^* + t) = \gamma(t^*) = \gamma_{t^*}$, where $\gamma_{t^*}$ is the constant exposure applicable in the interval $[t^*,t^* + t)$. 
The matrix form of the differential equation is then $$\boldsymbol{\bar{F}'}(t,t^*) =  \boldsymbol{\bar{F}}(t,t^*) (Q-\Lambda \gamma_{t^*}),$$ with solution $$
\boldsymbol{\bar{F}}(t,t^*) = \exp \left[ \left(Q-\Lambda \gamma_{t^*} \right) t \right].$$

This is derived using the following power series definition for exponential matrices:
\begin{align*}
\exp(\boldsymbol{A} t) = \boldsymbol{I} + t\boldsymbol{A} + \frac{t^2}{2!} \boldsymbol{A}^2 + \ldots,
\end{align*}
which holds when matrix $\boldsymbol{A}$ is commutative. In this case, this property is clearly satisifed as as $(Q-\Lambda \gamma_{t^*})$ is independent of $t$.  It follows using the above definition that $$\frac{d}{dt} \exp(\boldsymbol{A} t) =  \exp(\boldsymbol{A}t)\boldsymbol{A}.$$ Substitution of $\boldsymbol{A} = (Q-\Lambda \gamma_{t^*})$ and $\boldsymbol{\bar{F}}(t,t^*) = \exp \left[ \left(Q-\Lambda \gamma_{t^*} \right) t \right]$ then gives the required result.
\end{pfof}

\begin{remark}
	In the above theorem, we have set a constant value for the exposure measure in the interval $[t, t+t*)$. The reason for this approach is that without a constant exposure measure, the matrix differential equations are very difficult to solve. There are no known analytical solutions and while some approximations may be available through the use of Magnus expansions, these come at significant computational cost. It is for this reason that we restricted the exposure component to a function of bounded variation in Section \ref{sec:ModelSpec}, as any such function can be adequately approximated by the piecewise constant structure implied by the above theorem. We note that this is not unrealistically restrictive on the exposure function because if the variation in the function is large between two event arrivals, a middle point in the interval can be chosen to produce a better approximation. This approach is expanded upon later in this section.
\end{remark}

\begin{theorem}\label{thm:CKthm2}
	The probability of regime transitions with a single event arrival at time $t^* + t$, given the regime at time $t$, is expressed in matrix form as
	\begin{equation}
	\boldsymbol{f} (t,t^*)  = \exp \left[ \left(Q-\Lambda \gamma_{t^*} \right) t^* \right] \Lambda \gamma_{t+t^*}
	\end{equation}
	where $\gamma_{t+t^*}$ is the constant exposure measure applicable at time $t+t^*$.	
\end{theorem}
\begin{pfof}{Theorem \ref{thm:CKthm2}}
The formal definition of $f(t,t^*)$ is given by
\begin{align*}
f_{i,j} (t,t^*) & = \mathbb{P} [M(t^* + t)=j,N_M(\rho^{-1}(t^* + t^-))=n|M(t^*)=i,N_M(\rho^{-1}(t^*))=n] \nonumber \\
& \times \mathbb{P} [M(t^* + t) = j, N_M(\rho^{-1}(t^* + t))=n+1|M(t^* + t)=j,N_M(\rho^{-1}(t^* + t^-))=n]
\end{align*}
where $\rho(t^-) = \lim_{\epsilon \rightarrow 0} \rho(t-\epsilon)$. By following a similar set of derivations as in the proof for Theorem \ref{thm:CKthm1}, it can be seen that
\begin{align*}
\boldsymbol{f} (t,t^*) 	& = \exp \left[ \left(Q-\Lambda \gamma_{t^*} \right) t \right] \Lambda \gamma(t+t^*) \\ 
& = \exp \left[ \left(Q-\Lambda \gamma_{t^*} \right) t \right] \Lambda \gamma_{t+t^*}
\end{align*}
as required.
\end{pfof}

These theorems can also be intuitively understood as the non-homogeneous extensions of the corresponding MMPP expressions provided in \citet{Ryd96}. They provide the fundamental building blocks with which the EM algorithm for the MMNPP model is constructed. However, without further adjustments, use of these theorems would require the exposure measure to be unchanged within event inter-arrival time intervals. In situations where event occurrences are sparse, this may be a strong assumption. To remedy this issue, another event type is introduced for a change in the piecewise constant exposure function $\gamma$. For ease of notation, we rewrite the event times in chronological order. Denote all event times $t_k$  (the event of interest and changes in the exposure function). For each event time, there is an accompanying indicator $\delta_k$ to denote the type of event:
\begin{align*}
\delta_{k} = & \begin{cases}
0  ,& \text{if it is a change in the exposure function,} \\
1  ,& \text{if it is an event of interest}.
\end{cases}
\end{align*}
Applying this to Theorem \ref{thm:CKthm1}, the matrix function $\boldsymbol{f}$ becomes
\begin{align}
\boldsymbol{f}^{\delta_{k}}(t_{k-1},t_{k}-t_{k-1}) = & \begin{cases}
\boldsymbol{f}(t_{k-1},t_{k}-t_{k-1}) , &\delta_{k}=0,\\
\bar{\boldsymbol{F}}(t_{k-1},t_{k}-t_{k-1}) , &\delta_{k}=1.
\end{cases} = & \begin{cases}
\exp[(Q-\Lambda\gamma_{t_{k-1}}^{*})(t_{k}-t_{k-1})]\Lambda\gamma_{t_{k}}^{*} , & \delta_{k}=0,\\
\exp[(Q-\Lambda\gamma_{t_{k-1}}^{*})(t_{k}-t_{k-1})] , & \delta_{k}=1.
\end{cases} \label{eq:expchange}
\end{align}
The analytical expressions for the expectation of the E-step estimators are now written in terms of the above. Each estimator will be dealt with individually in the following subsections. Note that for clarity and readability, derivation of these estimators are provided in the appendix of the paper.

\subsubsection{The estimator for the number of regime changes: $\boldsymbol{\hat{a}_{i,j}}$}
The expression for the estimator of $m_{i,j}$ is given by
\begin{align}
\hat{a}_{i,j} 	& = \frac{q^0_{i,j}}{\boldsymbol{\pi}(Q^0,\Lambda^0) \bigg[ \prod_{k=1}^{n} \boldsymbol{f}^{\delta_k}(t_{k-1},t_k - t_{k-1}) \bigg] \boldsymbol{1}} \nonumber \\
& \times \Bigg[\int_{0}^{T} \boldsymbol{\pi}(Q^0,\Lambda^0) \bigg[ \prod_{k=1}^{N_M(\rho^{-1}(s)^-)} \boldsymbol{f}^{\delta_k}(t_{k-1},t_k - t_{k-1}) \bigg] \boldsymbol{\bar{F}} (t_{N_M(\rho^{-1}(s)^-)}, s - t_{N_M(\rho^{-1}(s)^-)}) \boldsymbol{1}_i \nonumber \\ 
& \times \boldsymbol{1}^\intercal_{j}  \boldsymbol{f}^{\delta_{N_M(\rho^{-1}(s)^-)+1}}(s,t_{N_M(\rho^{-1}(s)^-)+1} - s) \bigg[ \prod_{k=N_M(\rho^{-1}(s)^-)+2}^{n} \boldsymbol{f}^{\delta_k}(t_{k-1},t_k - t_{k-1}) \bigg] \boldsymbol{1} ds \Bigg] \label{eq:mhat}.
\end{align}

where $\boldsymbol{\pi}(Q^0,\Lambda^0 )$ is the starting distribution of state probabilities given the initial parameter estimates $Q^0$ and $\Lambda^0$, $\boldsymbol{1}$ is a $r\times 1$ vector of ones and $t_0 = 0$. Also, $\boldsymbol{1}_i$ is a vector of zeroes except for the $i$-th entry which is one. Finally, $k$ is now a counter for both event types and $n$ is the total number of original events and exposure changes.

\subsubsection{The estimator for the number of events of interest in each regime: $\boldsymbol{\hat{n}_{i}}$}

Continuing from Equation \eqref{eq:niestimator}, we obtain the following expression for the estimator of $n_i$:
\begin{align}\label{eq:niest2}
\hat{n}_{i}		& = \frac{1}{\boldsymbol{\pi}(Q^0,\Lambda^0) \bigg[ \prod_{k=1}^{n} \boldsymbol{f}^{\delta_k}(t_{k-1},t_k - t_{k-1}) \bigg] \boldsymbol{1}} \nonumber \\
& \times \sum_{k=1}^{n} \boldsymbol{\pi}(Q^0,\Lambda^0) \bigg[ \prod_{l=1}^{N_M(\rho^{-1}(t_k))} \boldsymbol{f}^{\delta_l}(t_{l-1},t_l - t_{l-1}) \bigg] \boldsymbol{1}_i \boldsymbol{1}_i^\intercal \bigg[ \prod_{l=N_M(\rho^{-1}(t_k)+1}^{n} \boldsymbol{f}^{\delta_l}(t_{l-1},t_l - t_{l-1}) \bigg] \boldsymbol{1}.
\end{align}

\subsubsection{The estimator for the time spent in each regime: $\boldsymbol{\hat{T}_{i}}$}
Using the expression for $\hat{T}_i$ from Equation \eqref{eq:tiestimator} and following a similar procedure to the $\hat{m}_{i,j}$ calculations, it can be seen that
\begin{align}
\hat{T}_i 	& = \frac{1}{\boldsymbol{\pi}(Q^0,\Lambda^0) \bigg[ \prod_{k=1}^{n} \boldsymbol{f}^{\delta_k}(t_{k-1},t_k - t_{k-1}) \bigg] \boldsymbol{1}} \nonumber \\
& \times \Bigg[\int_{0}^{T} \boldsymbol{\pi}(Q^0,\Lambda^0 ) \bigg[ \prod_{l=1}^{N_M(\rho^{-1}(s)^-)} \boldsymbol{f}^{\delta_l}(t_{l-1},t_l - t_{l-1}) \bigg] \boldsymbol{\bar{F}} (t_{N_M(\rho^{-1}(s)^-)},s - t_{N_M(\rho^{-1}(s)^-)}) \boldsymbol{1}_i \nonumber \\ 
& \times \boldsymbol{1}^\intercal_{j}  \boldsymbol{f}^{\delta_{N_M(\rho^{-1}(s)^-)+1}}(s,t_{N_M(\rho^{-1}(s)^-)+1} - s) \bigg[ \prod_{l=N_M(\rho^{-1}(s)^-)+2}^{n} \boldsymbol{f}^{\delta_l}(t_{l-1},t_l - t_{l-1}) \bigg] \boldsymbol{1} ds \Bigg]. \label{eq:That}
\end{align}
In the above, the event $\{M(s^-)=i\}$ has been replaced with $\{M(s)=i\}$ which does not change the density due to the fact that $\{M(s)\}$ is continuous in probability.

\subsubsection{The estimator for the operational time spent in each regime: $\boldsymbol{\hat{T}^*_{i}}$}
The results from the previous section are easily adapted to obtain the following estimator for $T^*_{i}$:
\begin{align}\label{eq:thatstarestim}
\hat{T}_i^* 	& = \int_{0}^{T} \mathbb{P} [M(s) = i | N_M(\rho^{-1}(t)), 0 \le t \le T ] \gamma(s) ds \nonumber\\
& = \frac{1}{\boldsymbol{\pi}(Q^0,\Lambda^0) \bigg[ \prod_{k=1}^{n} \boldsymbol{f}^{\delta_k}(t_{k-1},t_k - t_{k-1}) \bigg] \boldsymbol{1}} \nonumber \\
& \times \Bigg[\int_{0}^{T} \boldsymbol{\pi}(Q^0,\Lambda^0) \bigg[ \prod_{l=1}^{N_M(\rho^{-1}(s)^-)} \boldsymbol{f}^{\delta_l}(t_{l-1},t_l - t_{l-1}) \bigg] \boldsymbol{\bar{F}} (t_{N_M(\rho^{-1}(s)^-)}, s - t_{N_M(\rho^{-1}(s)^-)}) \boldsymbol{1}_i \nonumber \\ 
& \times \boldsymbol{1}^\intercal_{j}  \boldsymbol{f}^{\delta_{N_M(\rho^{-1}(s)^-)+1}} (s,t_{N_M(\rho^{-1}(s)^-)+1} - s) \bigg[ \prod_{l=N_M(\rho^{-1}(s)^-)+2}^{n} \boldsymbol{f}^{\delta_l}(t_{l-1},t_l - t_{l-1}) \bigg] \boldsymbol{1} \gamma(s) ds \Bigg]
\end{align}

\subsection{The scaled recursion algorithms in the adapted EM algorithm}\label{sec:ScaledRec}
The estimators derived in the previous section involve operations on the product of matrices, resulting in large computational cost. In the following section, we apply results from \citet{RoEpDi06} and \citet{Van78} to overcome the issues of computational cost and numerical instability. 

Firstly, some further notation is established for convenience. Let
\begin{align}
c_k 	& = \mathbb{P} [N_M(\rho^{-1}(t_k))= k, N_M(\rho^{-1}(t_k)^-)= k-1 | N_M(\rho^{-1}(t_{k-1}))=k-1, N_M(\rho^{-1}(t_{k-1})^-)=k-2, \nonumber \\   
& \quad N_M(\rho^{-1}(t_{k-2}))=k-2, N_M(\rho^{-1}(t_{k-2})^-)=k-3,\ldots],
\end{align}
be the conditional probability of observing the $k$-th event arrival at time $t_k$ given all previous event arrival times. Also, define $c_1 = \mathbb{P}[N_M(\rho^{-1}(t_1))= 1, N_M(\rho^{-1}(t_1)^-)= 0]$. An alternative formulation for the joint likelihood of observed event arrivals can be obtained using these $c_k$'s:
\begin{equation}\label{eq:ck}
\mathbb{P} [N_M(\rho^{-1}(t)), 0 \le t \le T] = \prod_{k=1}^{n} c_k.
\end{equation}
In a similar manner to the approach in \citet{RoEpDi06}, the forwards/backward recursion equations are defined by
\begin{align}
L(k) & = \boldsymbol{\pi}(Q^0,\Lambda^0) \prod_{l=1}^{k} \frac{\boldsymbol{f}^{\delta_l}(t_{l-1},t_l - t_{l-1})}{c_l}, \quad R(k) = \prod_{l=k}^{n} \frac{\boldsymbol{f}^{\delta_l}(t_{l-1},t_l - t_{l-1})}{c_l} \boldsymbol{1}, & k = 1,\ldots,n
\end{align}
with $L(0) = \boldsymbol{\pi}(Q^0,\Lambda^0)$ and $R(n+1) = \boldsymbol{1}$. Note that for each $k$, $L(k)$ is a $1\times r$ vector while $R(k)$ is a $r \times 1$ vector. Using the fact that
\begin{align}
c_k & = \boldsymbol{\pi}(Q^0,\Lambda^0 ) \prod_{l=1}^{k-1} \frac{\boldsymbol{f}^{\delta_l}(t_{l-1},t_l - t_{l-1})}{c_l} \boldsymbol{f}^{\delta_l}(t_{l-1},t_l - t_{l-1}) \boldsymbol{1} \nonumber \\ 
& = L(k-1) \boldsymbol{f}^{\delta_l}(t_{l-1},t_l - t_{l-1}) \boldsymbol{1},
\end{align}
the recursive functions can be iteratively computed as
\begin{align}
L(k) = \frac{L(k-1)\boldsymbol{f}^{\delta_k}(t_{k-1},t_k - t_{k-1})}{L(k-1)\boldsymbol{f}^{\delta_k}(t_{k-1},t_k - t_{k-1})\boldsymbol{1}}, \quad R(k) = \frac{\boldsymbol{f}^{\delta_k}(t_{k-1},t_k - t_{k-1}) R(k+1)}{L(k-1)\boldsymbol{f}^{\delta_k}(t_{k-1},t_k - t_{k-1})\boldsymbol{1}}.
\end{align}
The $L$ and $R$ functions can also be expressed in an alternative manner that is more interpretable. The $i$-th element in each vector is
\begin{align}
L(k)_{i} & = \mathbb{P}[M(t_k)=i | N_M(\rho^{-1}(t_{k-1}))=k-1, N_M(\rho^{-1}(t_{k-1})^-)=k-2,\ldots, N_M(\rho^{-1}(t_1))= 1, N_M(\rho^{-1}(t_1)^-)= 0], \label{eq:stateprob}\\
R(k)_{i} & = \frac{\mathbb{P}[N_M(\rho^{-1}(t_{n}))=n, N_M(\rho^{-1}(t_{n-1})^-)=n-1,\ldots,N_M(\rho^{-1}(t_{k}))=k, N_M(\rho^{-1}(t_{k})^-) = k-1 | M(t_{k-1})=i]}{B_k},\\
\text{with } B_k & = \mathbb{P}[N_M(\rho^{-1}(t_{n}))=n, N_M(\rho^{-1}(t_{n-1})^-)=n-1,\ldots,N_M(\rho^{-1}(t_{k}))=k, N_M(\rho^{-1}(t_{k})^-) = k-1 | \nonumber\\  
& \quad N_M(\rho^{-1}(t_{k-1}))=k-1, N_M(\rho^{-1}(t_{k-1})^-)=k-2,\ldots, N_M(\rho^{-1}(t_1))= 1, N_M(\rho^{-1}(t_1)^-)= 0]. \nonumber
\end{align}
The scaling procedure above overcomes the issue of numerical underflow when implementing this method on large data sets. It should also be noted that the derivations utilise the disjoint intervals of event/exposure change inter-arrivals where the exposure measure $\gamma(t)$ is constant so the results from Theorems \ref{thm:CKthm1}, \ref{thm:CKthm2} and Equation \eqref{eq:expchange} are applicable.

\subsubsection{The rescaled recursive algorithm for $\boldsymbol{\hat{m}_{i,j}}$} \label{sec:mhat}

In the following subsections, the scaled forwards/backwards recursions above will be applied to the E-step estimators derived in Section \ref{sec:Estep}. Beginning with the estimator for $\hat{m}_{i,j}$ obtained in Equation \eqref{eq:mhat}, factoring out the bracketed elements from the integrand and substituting the recursive functions $L$ and $R$ gives
\begin{align}
\hat{a}_{i,j} 	& = \sum_{k=1}^{n} \Bigg[ \frac{q^0_{i,j}}{c_k} \times \Bigg( \boldsymbol{\pi}(Q^0,\Lambda^0 ) \prod_{l=1}^{k-1} \frac{\boldsymbol{f}^{\delta_l}(t_{l-1},t_l - t_{l-1})}{c_l} \Bigg) \nonumber\\
& \times \int_{t_{k-1}}^{t_k^-} \boldsymbol{\bar{F}} (t_{k-1}, s - t_{k-1})) \boldsymbol{1}_i \boldsymbol{1}^\intercal_{j}  \boldsymbol{f}^{\delta_k}(s, t_{k} - s) ds \nonumber \\
& \times \Bigg( \prod_{l=k+1}^{n} \frac{\boldsymbol{f}^{\delta_l}(t_{l-1},t_l - t_{l-1})}{c_l} \boldsymbol{1} \Bigg) \Bigg]\\
& = q_{i,j}^0 \sum_{k=1}^{n}  \frac{1}{c_k} \int_{t_{k-1}}^{t_k^-} \boldsymbol{1}^\intercal_{j}  \boldsymbol{f}^{\delta_k}(s, t_{k} - s) R(k+1) L(k-1) \boldsymbol{\bar{F}} (t_{k-1}, s - t_{k-1})) \boldsymbol{1}_i ds
\end{align}
Using matrix notation, this becomes
\begin{align}
\boldsymbol{a}^\intercal 	& = \left(Q^{0}\right)^{\intercal} \odot \sum_{k=1}^{n}  \frac{1}{c_k} \int_{t_{k-1}}^{t_k^-} \boldsymbol{f}^{\delta_k}(s,t_{k} - s) R(k+1) L(k-1) \boldsymbol{\bar{F}} (t_{k-1}, s - t_{k-1}) ds \\ 
& = \left(Q \odot \sum_{k=1}^{n}  \frac{\mathcal{I}_k^\intercal}{c_k} \right)^\intercal,
\end{align}
where $\odot$ is the Hadamard product, otherwise known as the element-wise product for matrices. Also, the integral component above is represented by $\mathcal{I}_k$ where
\begin{align}
\mathcal{I}_k & = \int_{t_{k-1}}^{t_k^-} \boldsymbol{f}^{\delta_k}(s, t_{k} - s) R(k+1) L(k-1) \boldsymbol{\bar{F}} (t_{k-1}, s - t_{k-1}) ds \\
& = \begin{cases}
\int_{0}^{t_{k}^{-}-t_{k-1}}\exp[(Q-\Lambda\gamma_{t_{k-1}}^{*})(t_{k}-t_{k-1}-s)]\Lambda\gamma_{t_{k}}^* R(k+1)L(k-1)\exp[(Q-\Lambda\gamma_{t_{k-1}}^{*})(t_{k}-t_{k-1}-s)]ds ,& \delta_{k}=0,\\
\int_{0}^{t_{k}^{-}-t_{k-1}}\exp[(Q-\Lambda\gamma_{t_{k-1}}^{*})(t_{k}-t_{k-1}-s)]R(k+1)L(k-1)\exp[(Q-\Lambda\gamma_{t_{k-1}}^{*})(t_{k}-t_{k-1}-s)]ds, & \delta_{k}=1.
\end{cases}
\end{align}
where again, the integration is performed over disjoint time intervals where the exposure measure $\gamma(t)$ does not change. Thus, within each inter-arrival interval $\mathcal{I}_{k}$, the constant exposure measure will be denoted $\gamma_{k}$. The integrals within the cases above can be evaluated using the algorithm presented in \citet{Van78}. Define the following $2r \times 2r$ block-triangular matrices:
\begin{eqnarray}
C_{k}^{\delta_{k}} & = & \begin{cases}
\left[\begin{array}{cc}
Q-\Lambda\gamma_{t_{k-1}}^{*} & \Lambda\gamma_{t_{k}}R(k+1)L(k-1)\\
0 & Q-\Lambda\gamma_{t_{k-1}}^{*}
\end{array}\right] ,& \delta_{k}=0,  \\
\left[\begin{array}{cc}
Q-\Lambda\gamma_{t_{k-1}}^{*} & R(k+1)L(k-1)\\
0 & Q-\Lambda\gamma_{t_{k-1}}^{*}
\end{array}\right], & \delta_{k}=1.
\end{cases}
\end{eqnarray}
$\mathcal{I}_k$ can then be evaluated as the upper-right block of the matrix $\exp(C_k^{\delta_{k}} (t_k - t_{k-1}))$. This exponential can be easily and efficiently obtained using a variety of methods such as by using the diagonal Pad\'{e}'s approximant with repeated squaring \citep[see]{MoVa03}. Combining all of the above, the required estimator of $\boldsymbol{a}$ is obtained.

\subsubsection{The rescaled recursive algorithm for $\boldsymbol{\hat{n}_{i}}$}
Continuing on from Equation \ref{eq:niest2},
\begin{align}
\hat{n}_{i}	& = \sum_{k=1}^{n} \boldsymbol{\pi}(Q^0,\Lambda^0) \bigg[ \prod_{l=1}^{k} \frac{\boldsymbol{f}^{\delta_l}(t_{l-1},t_l - t_{l-1})}{c_l} \bigg] \boldsymbol{1}_{i} \boldsymbol{1}_{i}^\intercal \bigg[ \prod_{l=k+1}^{n} \frac{\boldsymbol{f}^{\delta_l}(t_{l-1},t_l - t_{l-1})}{c_l} \bigg] \boldsymbol{1} \nonumber \\
& = \big[ \sum_{k=1}^{n} L(k)^\intercal \odot R(k+1) \big]_{i}.
\end{align}
In matrix form, this is 
\begin{equation}
\boldsymbol{\hat{n}} = \sum_{k=1}^{n} L(k)^\intercal \odot R(k+1).
\end{equation}

\subsubsection{The rescaled recursive algorithm for $\boldsymbol{\hat{T}_{i}}$} 

The computation of $\hat{T}_i$ is very simple due to the following relationship between $\hat{m}_{i,i}$ and $\hat{T}_i$ (Equations \eqref{eq:mhat} and \eqref{eq:That} respectively):
\begin{align}
\hat{T}_{i} = \frac{\hat{m}_{i,i}}{q_{i,i}^0}.
\end{align}

\subsubsection{The rescaled recursive algorithm for $\boldsymbol{\hat{T}_{i}^*}$} 
The calculations for $\hat{T}_{i}^*$ are slightly more complex than for $\boldsymbol{\hat{T}_{i}}$. Continuing from Equation \ref{eq:thatstarestim},
\begin{align}
\hat{T}_i^* 	& = \sum_{k=1}^{n} \Big[ \frac{1}{c_{k}} L(k-1) \int_{t_{k-1}}^{t_k^-} \boldsymbol{\bar{F}} (t_{k-1}, s - t_{k-1})) \boldsymbol{1}_i \boldsymbol{1}^\intercal_{i}  \boldsymbol{f}^{\delta_k} (s, t_{k} - s) \gamma_k^* ds R(k+1) \Big].
\end{align}

These expressions were similar to the ones obtained when evaluating $\hat{a}_{i,i}$ in Section \ref{sec:mhat}, except that there is no $q_{i,j}$ constant in the summation and $i=j$ in this case. It is seen that
\begin{align}
\boldsymbol{T^*}	& = \text{diag} \left[ \sum_{k=1}^{n}  \frac{\gamma_k^* \mathcal{I}_k^\intercal}{c_k}\ \right]
\end{align}
where $\mathcal{I}_k$ has already been evaluated in the calculation of $\boldsymbol{\hat{a}}$ and the function diag denotes taking the diagonal entries of the matrix (as $\boldsymbol{\hat{T}^*}$ is a $r\times 1$ vector).

\subsection{The M-step of the EM algorithm}\label{sec:M-est}
In comparison to the E-step, the M-step of the EM algorithm is very simple. Under the constraint that $q_i = \sum_{j\ne i} q_{i,j}$, the unique maximum likelihood estimators of the MMNPP parameters are given in terms of the EM algorithm parameters by the following expressions: 
\begin{align}
\hat{q}_{i,j} = \frac{a_{i,j}}{T_{i}}, \quad \hat{\lambda}_{i} 	= \frac{n_{i}}{T^*_{i}}.
\end{align}
This is derived using the standard MLE procedure applied to the loglikelihood in Equation \eqref{eq:finalcomploglik}.

\subsection{Summary of adapted EM algorithm for MMNPPs} \label{sec:EMsum}
In summary, one iteration of the EM algorithm is given by the following:
\begin{enumerate}
	\item Using current estimates of $Q$ and $\Lambda$, evaluate $\boldsymbol{\hat{a}}, \boldsymbol{\hat{n}}, \boldsymbol{\hat{T}}$ and $\boldsymbol{\hat{T}^*}$.
	\item Using current estimates of $\boldsymbol{\hat{a}}, \boldsymbol{\hat{n}}, \boldsymbol{\hat{T}}$ and $\boldsymbol{\hat{T}^*}$, evaluate $Q$ and $\Lambda$.
\end{enumerate}

The last step is the determination of when the algorithm should be terminated. Two simple approaches may be applied. One method is to consider the change to the matrices $\boldsymbol{Q}$ and $\boldsymbol{\Lambda}$ and to stop the algorithm when the difference between successive EM iterations is smaller than some tolerance value. Another approach is to instead compare the value of the log-likelihood function, which is non-decreasing for successive iterations of the EM algorithm. From Equation \eqref{eq:ck} that the log-likelihood is given by
\begin{align}\label{eq:loglikeck}
\log & \,\mathbb{P} [N_M(\rho^{-1}(t_k))= k, N_M(\rho^{-1}(t_k)^-)= k-1, N_M(\rho^{-1}(t_{k-1}))=k-1, N_M(\rho^{-1}(t_{k-1})^-)=k-2,\nonumber \\   
& \quad N_M(\rho^{-1}(t_{k-2}))=k-2, N_M(\rho^{-1}(t_{k-2})^-)=k-3,\ldots, N_M(\rho^{-1}(t_1))= 1, N_M(\rho^{-1}(t_1)^-)= 0 ] \nonumber \\
& = \sum_{k=1}^{n} \log \mathbb{P} [N_M(\rho^{-1}(t_k))= k, N_M(\rho^{-1}(t_k)^-)= k-1 | N_M(\rho^{-1}(t_{k-1}))=k-1, N_M(\rho^{-1}(t_{k-1})^-)=k-2, \nonumber \\   
& \quad N_M(\rho^{-1}(t_{k-2}))=k-2, N_M(\rho^{-1}(t_{k-2})^-)=k-3,\ldots, N_M(\rho^{-1}(t_1))= 1, N_M(\rho^{-1}(t_1)^-)= 0], \nonumber \\
& = \sum_{k=1}^{n} \log c_k. 
\end{align}

Various simulation studies were carried out to ascertain the accuracy of the proposed algorithm. The results demonstrated a high level of accuracy and a number of other favorable attributes such as fast convergence of parameter estimates and robustness to initial inputs. For conciseness, we have only presented two case studies in Section \ref{sec:simcasestudy} the full details have been omitted from this paper but are available for interested readers in \citet{AvTaWoXi18}.

\subsection{Model order selection}\label{sec:Order selection}

An important aspect of the model is the selection of the order of the MMNPP. Some papers such as \citet{LaOzSo13} have proposed procedures for doing so although discussion of this aspect is relatively uncommon. Generally, simpler models such as order 2 or 3 are chosen for the purposes of demonstration or to maintain parsimony and interpretability of the underlying regimes.

From the previous section, the ease of obtaining the log-likelihood facilitates the use of likelihood-based model selection procedures. When choosing the optimal order of the MMNPP, the standard AIC and BIC criteria, as well as their extensions, are straightforward to evaluate. However, we suggest the usage of a more data-driven approach to order selection.

Our proposed methodology for order selection in the absence of (or to complement) expert opinion is as follows. For a chosen order $x$ (starting at order 2, provided there is evidence of regimes as discussed later in Section \ref{sec:evidenceregimes}), residuals are calculated as the difference between the observed number of events and the predicted number of events. A white-noise test then applied, and if the evidence of the white-noise property is rejected, then the MMNPP model of order $x+1$ is fitted instead and the process iterates until a suitable $p$-value for the hypothesis of white noise residuals is obtained. In the empirical case study shown in Section \ref{sec:CaseStudy}, the Bartlett B test for white noise is used \citep[see][]{Bar67}.

The methodology outlined above for order selection is an intuitive approach based on our implementation and interpretation of the modelling framework. This framework seeks to capture as much modellable/known information as possible in the exposure/volume measure $\gamma$. The hidden Markov layer then captures any residual data features that may be left over such as auto-correlation and over-dispersion. From this perspective, a model selection criterion based on white-noise residual testing is very natural as it means that there are a sufficient number of regimes in the hidden Markov layer to capture the leftover residual information after the data has been processed by the $\gamma$ component. Thus, this methodology sets a minimal number of regimes required to remove data characteristics like auto-correlation. 

Further questioning of whether an even higher number of states is necessary can be done through standard procedures such as the Akaike/Bayesian information criteria or by more qualitative assessments based on the importance of interpreting the underlying regimes. However, we note that the relationship between likelihoods and the presence of data features such as auto-correlation is less clear compared to examinations of residuals. Thus, we believe that our proposed white-noise residual testing methodology leads to a well-fitted model that balances parsimony and interpretability.

\section{Numerical illustration: Simulated case study}\label{sec:simcasestudy}
In this section, we include a short simulation case study to demonstrate the accuracy and efficiency of the calibration procedure we have detailed in the previous section. It is seen that the algorithm produces estimates that are close to the true values, and this convergence occurs over a reasonably large range of starting parameter estimates for the algorithm. Further, the computational advantages described in the previous sections are highlighted as calibrations over tens of thousands of events required only a few minutes, which is sufficiently fast for practical implementation.

In this case study, we simulate events from a defined MMNPP in order to compare estimates with their true values. For simplicity, an order 3 MMNPP was chosen. The chosen parameter values for the generator matrix of the Markov chain and the baseline Poisson intensities are given respectively by

\begin{equation}
Q=\left[\begin{array}{ccc}
-0.8 	& 0.5 	& 0.3 \\
0.6 	& -1	& 0.4 \\
0.3		& 0.5	& -0.8 \\
\end{array}\right], \,\,\,\lambda=\left(5,10,20\right).
\end{equation}

Finally, the exposure function $\gamma$ was a piecewise function that varied periodically. For example, the simulation below varied the exposure function every 100 time units. It is noted that the frequency of this change was tested at a variety of different levels from 10 to 200 time units and it is shown to have negligible impact on the parameter estimates for $Q$ and $\Lambda$.

Events are simulated over a period of 1,000 time units which produces tens of thousands of event arrival times. In all simulations, the total calibration time using the EM algorithm took less than 5 minutes.

The calibrated parameter estimates for $Q$ and $\Lambda$ are

\begin{equation}
\hat{Q}=\left[\begin{array}{ccc}
-0.77 		& 0.48 			& 0.29 \\
0.62 		& -0.97			& 0.35 \\
0.27		& 0.54			& -0.81 \\
\end{array}\right], \,\,\,\hat{\lambda}=\left(5.07, 10.09, 20.37\right).
\end{equation}

It can be seen that the calibrated values are all quite close to the true values. In particular, the frequencies associated with each regime are very precise. These results are consistent across many simulation specifications. Also, the calibrated values demonstrate some robustness as convergence to the true values was maintained over a relatively large range of initial estimates.

\subsection{Calibration times}
As previously mentioned, the calibration times in this simulation case study of approximately 20,000 events took less than 10 minutes, which is very quick for such volumes of data. In addition, our empirical case study detailed in Section \ref{sec:CaseStudy} contains a calibration involving almost three-quarters of a million claims, and that calibration took approximately 4 hours to complete. As a result, we believe that we have sufficiently reduced the calibration times to practicably implementable levels.

We conducted another small simulation study to examine the calibration times and memory requirements associated with different choices of MMNPP order as well as different data set sizes. This study was performed on a desktop computer with an i7-4790 CPU @ 3.60GHz and the results are shown in Table \ref{table:simcompstudy} below.

\begin{table}[H]
	\begin{tabular}{|l|l|r|r|r|r|r|r|}
		\hline
		{ }                  & { \textbf{Num Obs}} & { \textbf{1,000}} & { \textbf{2,000}} & { \textbf{5,000}} & { \textbf{10,000}} & { \textbf{100,000}} & { \textbf{500,000}} \\ \hline
		{ \textbf{Order 2}}  & { Timing (sec)}     & { 21.79}          & { 42.17}          & { 107.09}         & { 210.46}          & { 1816.7}           & { 7517.73}          \\ \hline
		{ }                  & { RAM Usage (MB)}   & { 0.32}           & { 0.49}           & { 1.12}           & { 2.2}             & { 21.46}            & { 96.87}            \\ \hline
		{ \textbf{Order 10}} & { Timing (sec)}     & { 59.03}          & { 118.30}         & { 334.62}         & { 549.50}          & { 4783.77}          & { 24372.67}         \\ \hline
		{ }                  & { RAM Usage (MB)}   & { 4.12}           & { 8.1}            & { 20.07}          & { 40.06}           & { 399.86}           & { 1990.98}          \\ \hline
	\end{tabular}
	\caption{Timing and memory usage by number of observations and order of the MMNPP}	\label{table:simcompstudy}
\end{table}

We can see that the computation times required grow proportionately with the size of the data set that the parameters must be calibrated on. However, when moving from an order 2 to and order 10 model, the increase in computation times is much smaller despite the increasing number of parameters that need to be estimated (going from order $r$ to order $r+1$ requires an extra $2r + 1$ number of parameters).

In terms of the memory requirements, it is seen in the above table that for even very large data sets of half a million observations and an extreme case of an order 10 MMNPP, the RAM usage of the algorithm is still within feasible ranges of most computers. However, we expect that with most implementations of the proposed MMNPP framework, it is preferable to maintain a lower number of regimes for reasons of parsimony and interpretability. As a result, we believe that the proposed modelling framework is applicable to a variety of modelling procedures involving large data sets.

\subsection{Convergence rates}
We conclude the simulated case study with a note on the convergence of the parameters. When the iterative EM calibration algorithm is terminated using a likelihood condition, it is beneficial if the convergence to the maximal likelihood value is suitably regular. This is indeed the case, as can be seen in Figure \ref{fig:loglikeconv} below. 

\begin{figure}[H]
	\centering
	\includegraphics[width=\linewidth,trim={0 0.5cm 0 2cm},clip]{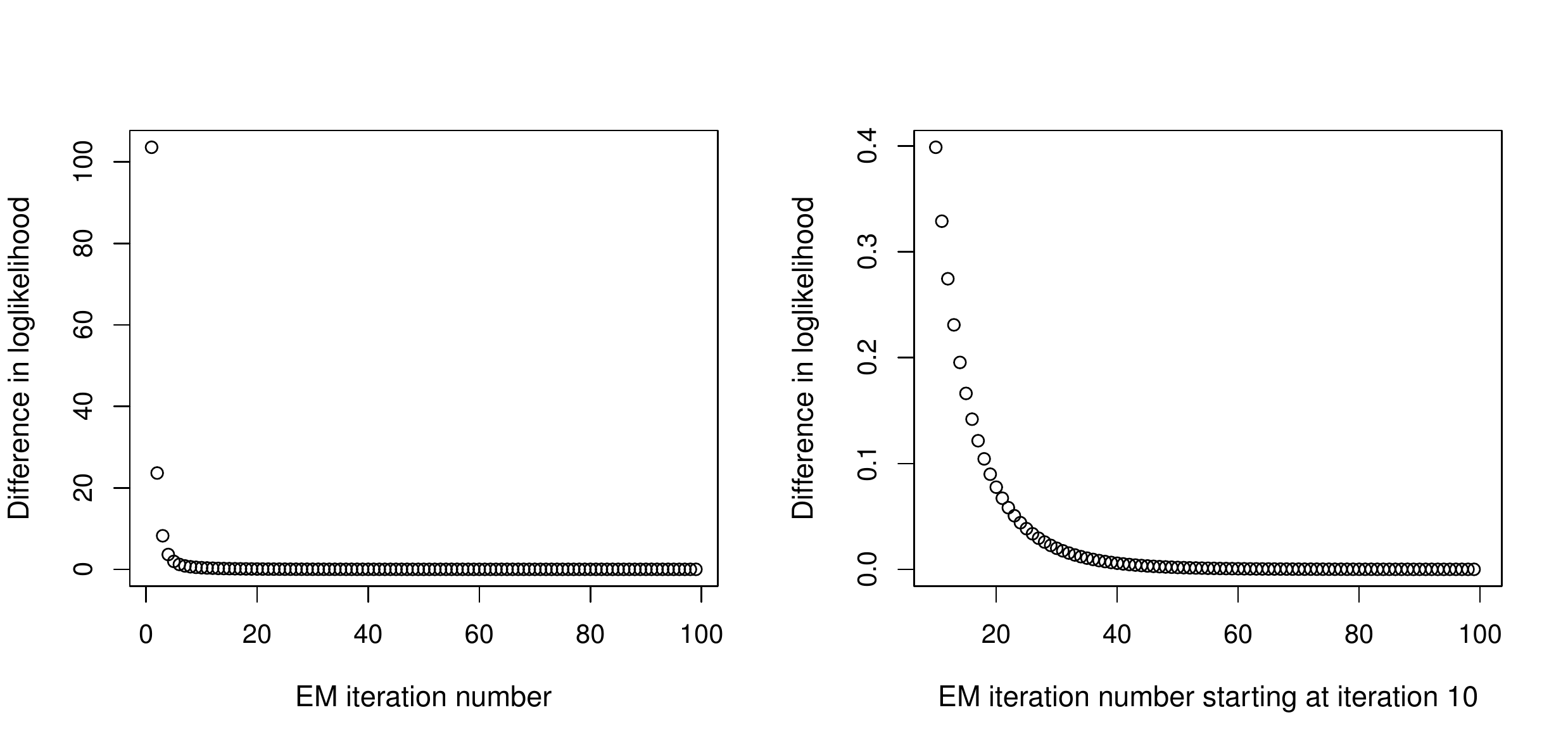}
	\caption{Plot of the difference between successive loglikelihood values} \label{fig:loglikeconv}
\end{figure}

It can be seen that the loglikelihood converges to the maximal value in a very fast, exponential-like fashion. This demonstrates the beneficial convergence properties of the proposed calibration algorithm. It can be seen that this type of convergence holds not only for the loglikelihood but also for the parameter estimates as well. Figure \ref{fig:paramconv} below provides an example of a similar quick convergence rate for the frequency and transition parameters, despite initial estimates being somewhat far from their true values:

\begin{figure}[H]
	\centering
	\includegraphics[width=1\linewidth,trim={0 0.5cm 0 2cm},clip]{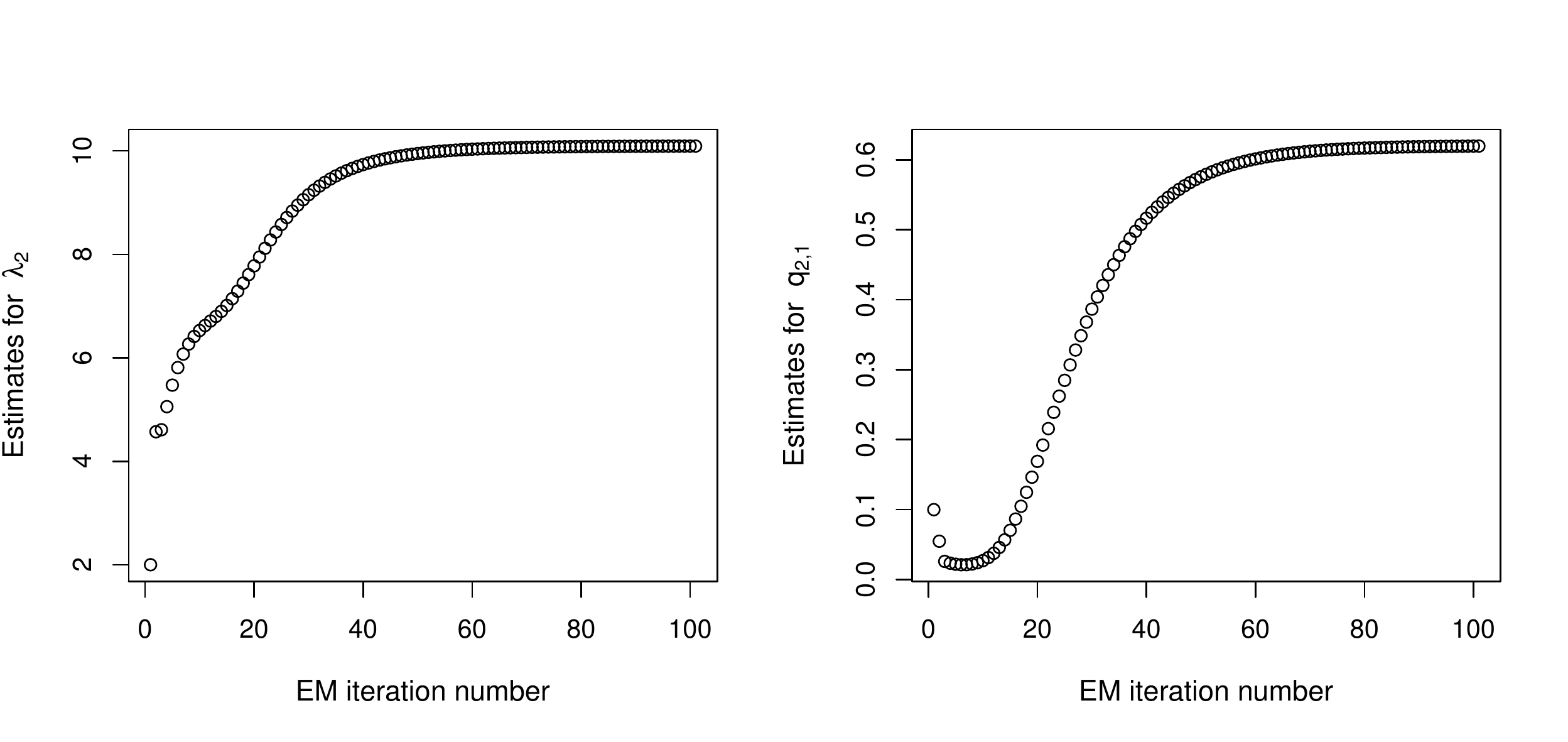}
	\caption{Plot of successive parameter estimates for $\lambda_2$ and $q_{2,1}$} \label{fig:paramconv}
\end{figure}

From comparison of the convergence plots, we suggest a stopping criterion based on a likelihood difference of $1\times 10^{-4}$, which results in about 60 iterations of the EM algorithm in this case. 

\section{Numerical illustration: Motor insurance claims} \label{sec:CaseStudy}
In the following section, we apply the MMNPP model to a set of motor insurance real data to determine what insights can be gained in order to assist insurance operations in a real world setting; see also Section \ref{S_ins} for additional background. Insurance data sets have several characteristics that call for the additional features introduced in this paper. These data sets generally consists of a large number of event observations which, as seen in the previous section, facilities more accurate detection of the underlying regimes. Further, the modelling of claims in motor lines of business is standard, which assists in the analysis required to determine the appropriate exposure measure $\gamma$. How we decided which trends to include in $\gamma$ is outlined in the Section \ref{sec:VolumeJust}. The residual components are modelled by a MMNPP. Additional insights available through the MMNPP model are discussed.

\subsection{The AUSI data set}
The Allianz, University of New South Wales, Suncorp and Insurance Australia Group (AUSI) dataset was developed as part of an Australian Research Council Linkage Project. The data set contains non-life insurance claims records in a standard format containing information on each claim transaction such as occurrence and notification dates, claim states and finalisation dates. It also includes a policy file which describes underwritten policies, containing information such as date of inception/expiry and sum insured. 

This case study investigates the Private Motor line of business from the 1st of January, 2005 to the 31st of December, 2010. It is noted that the data set extends to the end of 2014 and due to the short-tailed nature of the reporting delay for motor claims, all accidents that occurred during this period are expected to have been reported in the data set. As a result, considerations of censoring and reporting delay are not material here. Further, all claims related to catastrophe events have been removed from this data set. This is due to the fact that catastrophic claims generally exhibit very different characteristics compared to standard claims and thus it is standard practice to separate their analysis from the main body of the data. Otherwise, their presence may distort the analysis (\citet{HaBuHo96}). In total, the utilized subset of the data consists of significantly more than half a million claim observations. 

Because the level of granularity of the claims data is not finer than daily, an extra assumption is required to distribute the claim arrival times throughout the day, as the MMNPP is a continuous time model. In order to not introduce any artificial variability, the claims are assumed to arrive at equal intervals throughout the day. This approach was compared using simulation studies against simulating claim arrivals using a uniform distribution but the latter approach induced a small amount of additional variability, clouding results in terms of regime identification. Further, issues arose in terms of the appropriate method to aggregate these simulations. Finally, the use of simulations meant that the model had to be fit several times, increasing the computational burden of the approach. Thus, the more convenient approach of an equidistant distribution was taken instead.

\subsection{Preliminary data analysis}
We present some figures in the following section to provide the reader some sense of the data. Due to confidentiality, identifying features such as y-axes of plots have been removed.

\begin{figure}[H]
	\centering
	\includegraphics[width=1\linewidth,trim={0 0.5cm 0 2cm},clip]{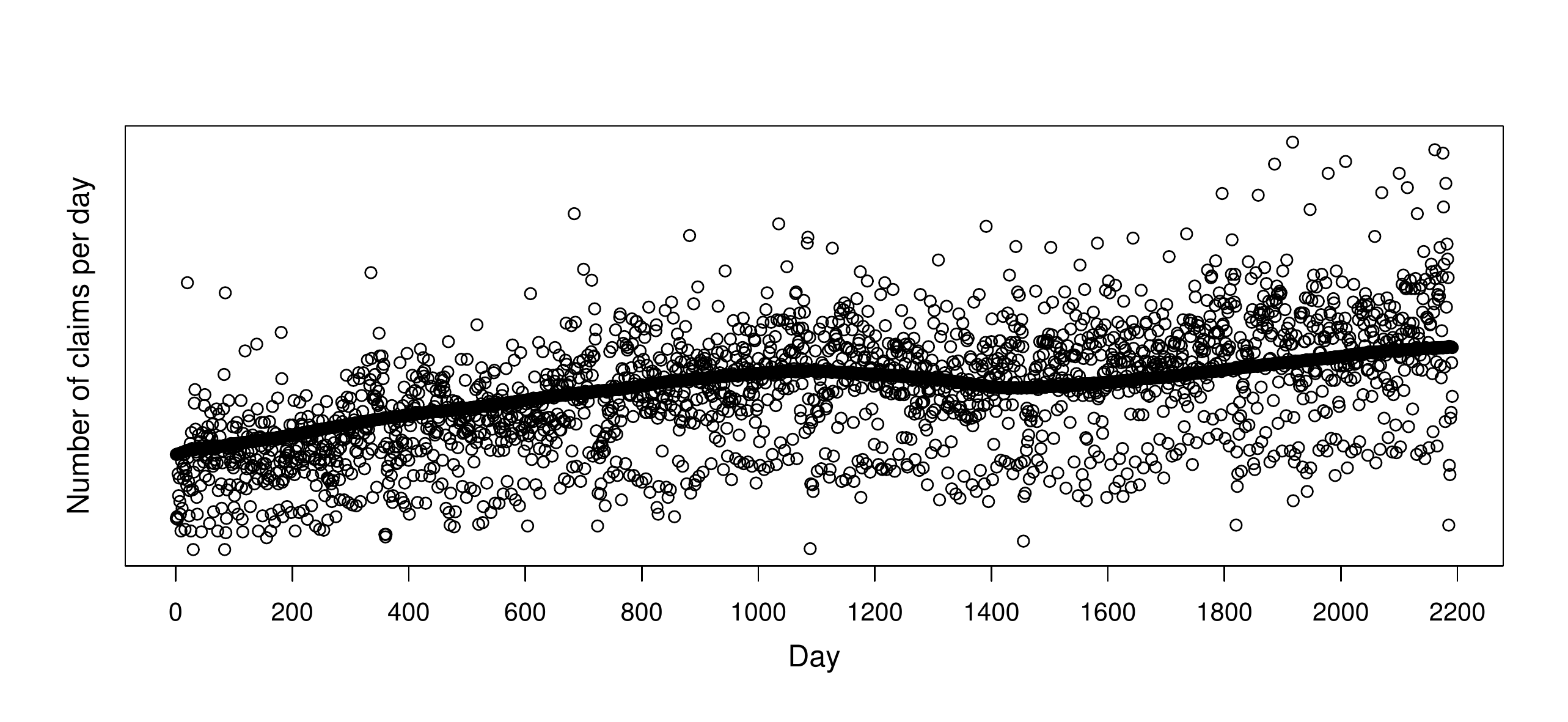}
	\caption{Number of claims and policies in force by day} \label{fig:prelimdata}
\end{figure}

Figure \ref{fig:prelimdata} below plots the evolution of the number of claims per day over the investigated period of 6 years while the line depicts the number of insurance policies in force over the same period.

\begin{figure}[H]
	\centering
	\includegraphics[width=1\linewidth,trim={0 0.5cm 0 2cm},clip]{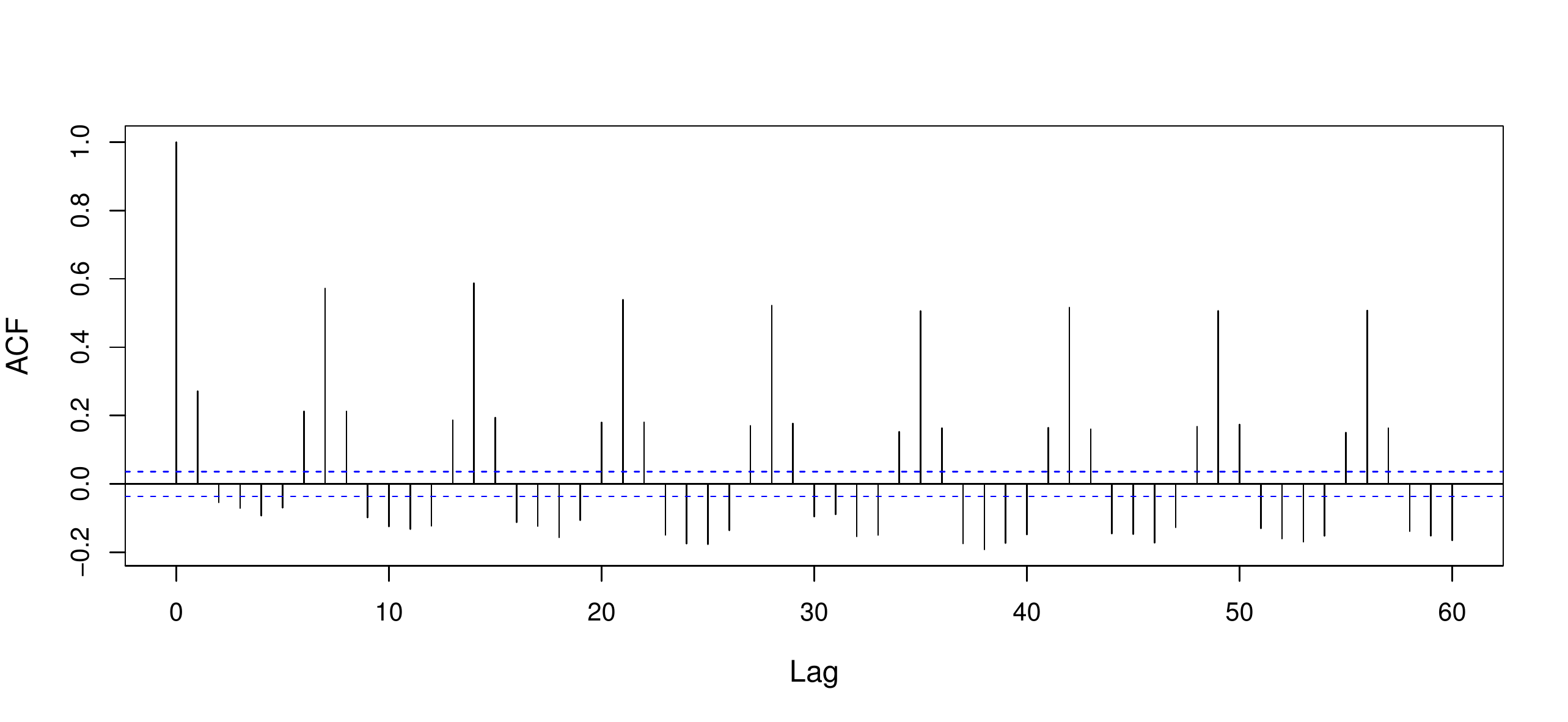}
	\caption{Plot of the auto-correlation function of daily claim events scaled by the number of policies} \label{fig:prelimauto}
\end{figure}

We examine the daily claim frequencies are scaled by the number of insurance policies in force. Figure \ref{fig:prelimauto} below plots the auto-correlation function up to a lag of 60 days and some data features are evident, such as the strong weekly seasonality within the data. This is not unexpected and known data features such as these are incorporated into the volume/exposure component and details of how this procedure is applied is given in the next section.

\subsection{Calibration of the volume/exposure component}\label{sec:VolumeJust}
Within the context of motor claims analysis, several factors are typically incorporated into the frequency perturbation measure $\gamma$. Clearly, the number of policies in force are a relevant exposure component but other trends such as seasonality also exist. An over-dispersed Poisson (ODP) GLM was chosen as an appropriate model to determine the exposure process $\gamma$. Such an approach is standard in actuarial modelling. Covariates of interest were identified using a combination of data analysis and domain knowledge. Statistical significance was then tested and the final model consisted of the features relating to the number of policies in force (introduced as an offset), various forms of seasonality such as weekday/weekend, public holiday, month and day of month fluctuations, and a linear residual component to capture the gradual decay of motor claim frequencies over time.

Further analysis was conducted of $\gamma$ over time and no additional events involving sudden changes in the exposure component were deemed necessary. This is partly due to the large number of claim events within the data set.

\subsubsection{Evidence of hidden regimes}\label{sec:evidenceregimes}
A natural question that may arise is whether hidden regimes exist within the data after having accounted for the data features above. A number of statistical results indicate that this is indeed the case. Firstly, the dispersion parameter calibrated in the ODP GLM was 3.16, indicating some level of over-dispersion within the data relative to the standard Poisson process. 

Further, a Wald-Wolfowitz runs test was also applied in order to test for persistence of the frequency process within regimes. This procedure tested for randomness in the runs above and below the Poisson estimate of the intensity, having adjusted for the factors in the volume component outlined in the previous subsection. The obtained p-value was 0, suggesting that the frequencies tended to remain at certain levels over time.

A white-noise residual testing procedure was utilised for order selection, and an order 4 MMNPP was identified as optimal; see Section \ref{sec:Order selection} for methodological details.

Finally, we produce another auto-correlation plot using the residuals of the ODP GLM fit to test for serial correlation features that are not yet taken into account by the GLM. Indeed, as the optimal order of the MMNPP model is 4 (and not 1), it is expected that there still exists some residual auto-correlation left in the time series data and this is corroborated in Figure \ref{fig:GLMauto} below.

\begin{figure}[H]
	\centering
	\includegraphics[width=1\linewidth,trim={0 0.5cm 0 2cm},clip]{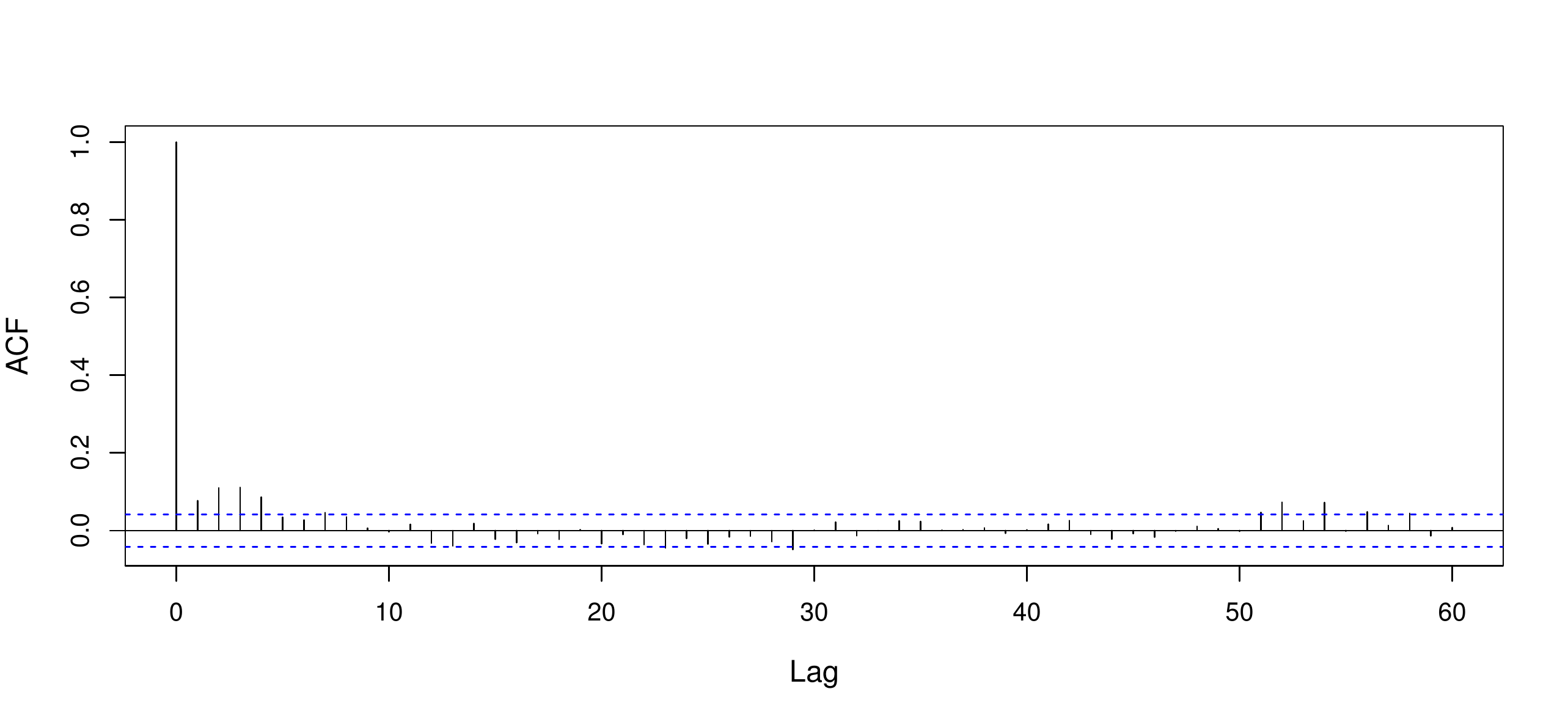}
	\caption{Plot of the auto-correlation function of the ODP GLM residuals} \label{fig:GLMauto}
\end{figure}

\subsection{Calibration of the hidden Markov component}
After obtaining the exposure component, we now move to the calibration of the hidden Markov component. As previously mentioned, an order 4 MMNPP was chosen and the final calibrated parameters were
\begin{equation}
Q=\left[\begin{array}{cccc}
-0.39 	& 0.08 	& 0.28  & 0.02 \\
0.00 	& -0.06	& 0.05  & 0.00 \\
0.38	& 0.05	& -0.43 & 0.00 \\
1.00	& 0.00	& 0.00 & -1.00 \\
\end{array}\right], \quad \lambda=\left(135,177,204,518\right).
\end{equation}
The distance between Poisson claim intensities suggest that the identified regimes are distinct. The generator matrix provides information about to the transition probabilities between regimes averaged over the entire period and modellers can also obtain other inferences such as how many claims occurred within each regime and how much time was spent in each regime through the estimators used in the EM algorithm. This is displayed in Table \ref{table:casestudymest}. For confidentiality reasons, the number of claims in each regime statistic has been provided as a proportion of total claims instead.

\begin{table}[H]
	\centering
	\begin{tabular}{|l|l|}
		\hline
		\textbf{Statistic of interest}                        & \textbf{Estimator} \\ \hline
		Number of changes to each regime & $\left[\begin{array}{cccc}
		-20.2 	& 5.9 		& 12.4 	& 1.0 \\
		4.4 	& -106.1	& 101.3 & 0.0 \\
		98.9	& 14.8		& -113.0& 0.0 \\
		1.0		& 0.0		& 0.0	& -1.0 \\ 
		\end{array}\right]$          \\ \hline
		Proportion of claims in each regime  &   (2.3\%, 88.9\%, 8.7\%, 0.0\%)        \\ \hline
		Time spent in each regime (days)        &   (52.4, 1870.5, 267.1, 1.0)        \\ \hline
	\end{tabular}
	\caption{E-estimator quantities of interest}
	\label{table:casestudymest}
\end{table}

\begin{figure}[H]
	\centering
	\includegraphics[width=1\linewidth,trim={0 0.3cm 0 2.1cm},clip]{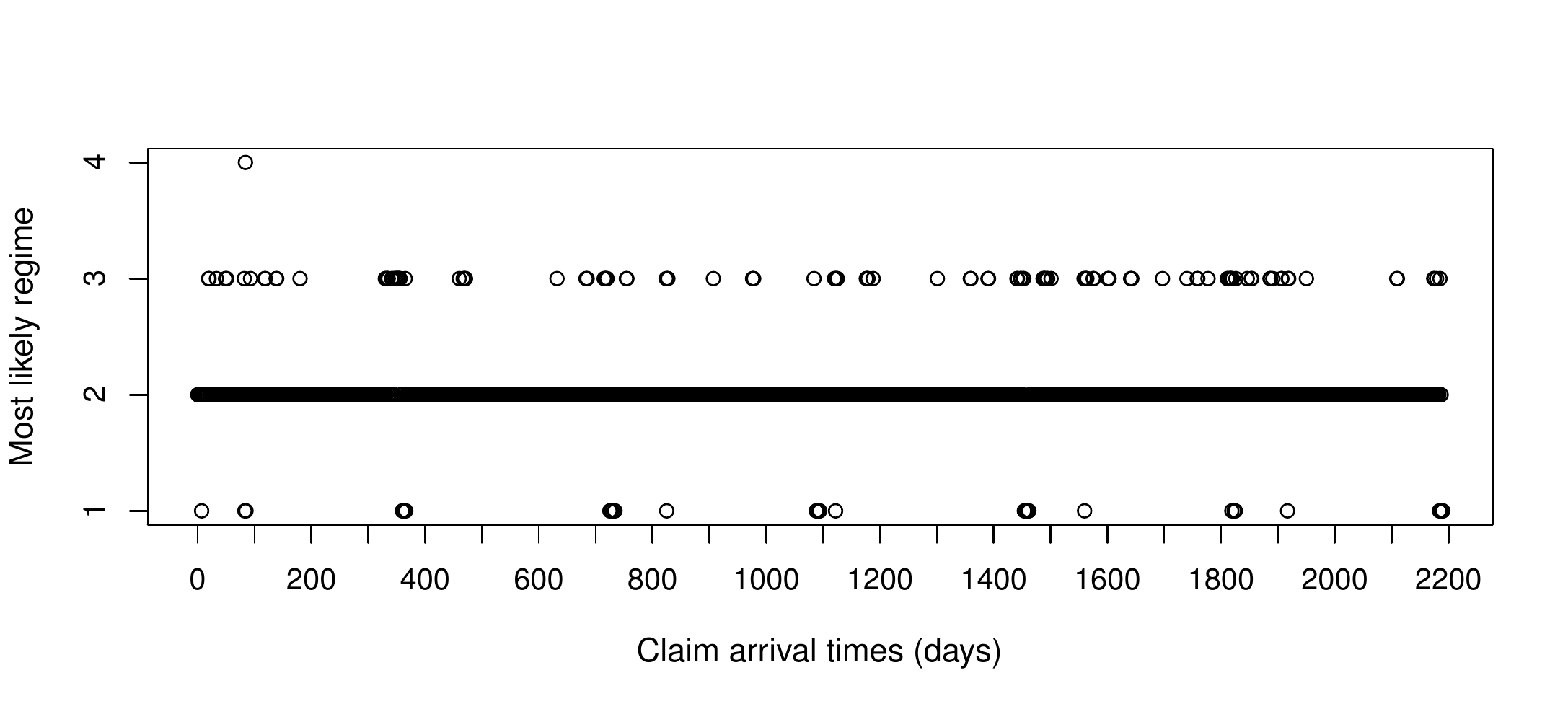}
	\caption{Most likely regimes over time} \label{fig:casestudyreg}
\end{figure}

Furthermore, Equation \eqref{eq:loglikeck} allows for the generation of Figure \ref{fig:casestudyreg}, which depicts the most likely regime at each claim arrival time. Several insights are able to be drawn here. Firstly, there is a single day (a public holiday coinciding with a weekend) with an extremely high number of claims compared to expectations, sufficiently so to warrant a separate regime (regime 4). As insurance claims related to catastrophic events have been removed, such an explanation is unlikely.  This result demonstrates some robustness of the MMNPP model to outliers, as the observation has been objectively identified and its impact on the other observations and regimes has been mitigated. This mitigation effect is due to the fact that the high intensity associated with the fourth regime will not skew the parameter estimates for the other regimes as significantly. For example, if the order of the MMNPP model was instead three, then this outlying observation would have increased the frequency parameter associated with the highest regime.

The remaining sequence of most likely regimes seems readily interpretable. There is a ``normal" state (State 2) within which the process spends the majority of the investigation period. There are occasional jumps to the higher frequency regime (State 3) and somewhat periodic jumps to the lower frequency regime (State 1). This latter result produces some interesting insights which we discuss in the next section.

\subsection{Utilising the MMNPP model for model development}\label{sec:Modeldev}
A secondary observation demonstrates the diagnostic property of the MMNPP model. Figure \ref{fig:casestudyreg} suggests that the adjusted frequency process remains in State 2 with occasional jumps to a higher or lower state. A closer look leads to the identification of  periodic transitions to State 1. This appears to be due to the effect of workers taking leave from the 27th of December until the new year. This effect was not originally captured in the known component $\gamma$ but is fairly intuitive in hindsight. The reduction in claims on the 25th and 26th was already captured (due to these dates being public holidays), but to further include the following dates until New Year's Eve was not obvious, and could easily be challenged in absence of evidence such as provided by our methodology.

Modellers may encounter difficulties in properly identifying all relevant factors of interest for analysis in practice. In the example above, upon discovery of the ``end-of-year" effect on claim frequencies, this covariate can be included within the $\gamma$ component of the model. This iterative exercise produces model improvement as the more interpretable volume/exposure component is expanded. Further, the impact of any additionally identified factors are removed from the hidden Markov chain, allowing this component to more realistically represent data features that are truly unobservable to the modeller. Repeated application of this diagnostic feature of the MMNPP model is facilitated by the computational speed increases produced in the previous sections. The final result is ideally a model where all known features are accounted for through the frequency perturbation measure $\gamma$ while the hidden Markov regimes capture the aggregated effects of factors that are unable to be explicitly modelled.

\subsection{Model comparisons}
In order to demonstrate the benefits provided by the MMNPP approach as opposed to more elementary ones, we now provide a comparison between the MMNPP model and four special (simpler) cases, namely: (1) a homogeneous Poisson process (which serves as the null model), (2) a non-homogeneous Poisson process (with the non-homogeneous component incorporating the effect of covariates through the ODP GLM), (3) a homogeneous Markov-modulated Poisson process with 3 states, and (4) another homogeneous MMPP model with 10 states.

Model (2) extends Model (1) by allowing a non-homogeneous (but deterministic) intensity, whereas Model (3) extends Model (1) by making the constant intensity regime-dependent. However, it can be seen below that the predictions produced by Model (3) are suboptimal and thus Model (4) is introduced to extend Model (3) through the use of a greater number of regimes. The MMNPP considered in this paper combines the extensions outlined above, and the comparisons below suggest that both are necessary for the data set under consideration. 

\begin{remark}\label{rem:order3MMNPP}
We note here that after identification of the outlying regime and end of year effects discussed in the previous sections, these covariates were included in the ODP GLM associated with the $\gamma$ frequency perturbation component. The order selection methodology described in Section \ref{sec:Order selection} was again applied, resulting in a 3 regime MMNPP being chosen as optimal. Analysis of model outputs shows that essentially, the outlying regime (regime 4) has been removed while the lower regime (regime 1) has been retained as there are still a sufficient number of events to warrant a separate regime. Thus, we emphasise that in these comparisons, the chosen ODP GLM incorporates the additional features described in Section \ref{sec:Modeldev}. This is generous, because without use of the MMNPP framework, some of these data features may not have been identified.
\end{remark}

\begin{remark}
The chosen number of states/regimes in the MMPP models is also based on the residual-testing methodology outlined previously. The order 3 MMPP model was chosen to match the MMNPP model. Model (4) attempted to utilise a similar residual-testing method to obtain the optimal number of states for an MMPP model. However, even upon reaching 10 states, the residuals still did not resemble white noise. A higher number of states would continue to improve in-sample fit but this comes at the cost of reduced model interpretability. Thus, our chosen number of states is 10 in this case.
\end{remark}

\begin{figure}[htb]
	\centering
	\includegraphics[width=1\linewidth,trim={0 0cm 0 0cm},clip]{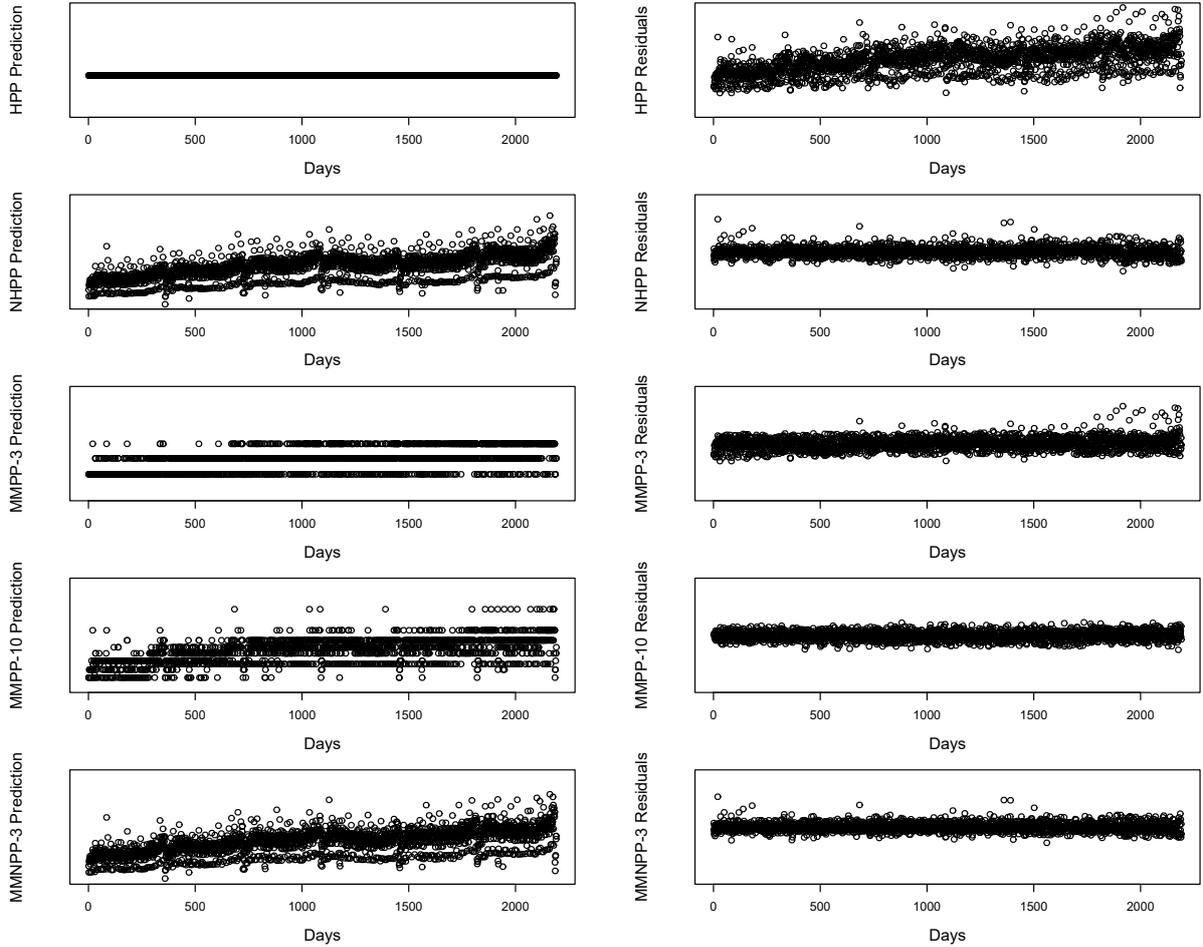}
	\caption{Predictions and residuals for each model} \label{fig:MMNPPcomp}
\end{figure}

For each of the models, we produce daily in-sample predictions and then examine the residuals. The daily granularity was chosen to match the granularity of the original data. Figure \ref{fig:MMNPPcomp} plots the predictions in the left column and the residuals for each model in the right column. Note that for confidentiality reasons, the y-axis has been removed.

Table \ref{table:residualstats} provides summary statistics of the residuals for each model, as well as the $p$-value for the Ljung-Box tests for auto-correlation up to lags of 91, 182 and 365 days and the Wolf-Waldowitz runs test for randomness/persistence. These lags were chosen as they are the prevalent granularities at which insurance data is examined. Finally, the last row gives an indication of the calibration time involved for each model.

\begin{table}[htb]
	\centering
\begin{tabular}{l|l|l|l|l|l|}
	\cline{2-6}
	& HPP        & NHPP      & MMPP-3    & MMPP-10   & MMNPP-3   \\ \hline
	\multicolumn{1}{|l|}{$\Sigma$ residuals}            & 0.0        & 0.0       & 748.8     & -1130     & 1418      \\ \hline
	\multicolumn{1}{|l|}{$\Sigma$ absolute residuals}   & 130,644    & 45,211    & 59,501    & 44,487    & 40,203    \\ \hline
	\multicolumn{1}{|l|}{$\Sigma$ residuals squared}    & 12,051,219 & 1,670,274 & 2,725,133 & 1,385,661 & 1,295,155 \\ \hline
	\multicolumn{1}{|l|}{Ljung-Box test p-value (lag = 91)} & 0          & 8e-08     & 0         & 0         & 0.08      \\ \hline
	\multicolumn{1}{|l|}{Ljung-Box test p-value (lag = 181)} & 0          & 1e-06     & 0         & 0         & 0.21      \\ \hline
	\multicolumn{1}{|l|}{Ljung-Box test p-value (lag = 365)} & 0          & 2e-06     & 0         & 0         & 0.12      \\ \hline
	\multicolumn{1}{|l|}{Runs test p-value}      & 0          & 9e-08     & 1e-05     & 0.96      & 0.85      \\ \hline
	\multicolumn{1}{|l|}{Calibration time}      & $<$1 min          & $<$1 min     & $\sim$4 hours     & $\sim$6 hours      & $\sim$4 hours      \\ \hline
\end{tabular}
	\caption{Summary statistics for model comparison}
	\label{table:residualstats}
\end{table}

It is clear from the results that the homogeneous Poisson prediction (HPP) produces a poor fit. Further, all residual summary statistics are significantly greater than the ones for any other considered model.

Model NHPP (2) uses the same over-dispersed Poisson GLM for the non-homogeneous intensity function as is used for the MMNPP-3 model. The aim of this approach is to incorporate domain knowledge within the model-building process. In addition, this method is standard in the industry both due to ease of implementation as well as model understandability and tractability. From Table \ref{table:residualstats}, this approach provides a marked improvement upon Model (1). The residual plot also seem more reasonable in comparison. However, both statistical tests applied to Model NHPP (2) indicate that the residuals do not resemble white noise and thus, there remain some data features that have still not been adequately accounted for. 

Model MMPP-3 (3) attempts to perform the same task as Model NHPP (2) through the use of the more-objective Markov-modulated Poisson process. However, it can be seen from the prediction plots of Models NHPP (2), MMPP-10 (4) and the MMNPP-3 that there is a increasing trend in the raw claims frequency over time. The prediction plot for Model MMPP-3 (3) illustrates that with only 3 regimes, the MMPP model provides a poor fit to the data. Evidence of inferior fit compared to other models under consideration is also seen in Table \ref{table:residualstats}.

The 10-state MMPP approach (4) is a more data-driven approach compared to Model NHPP (2), and the high number of regimes means that the fit is able to better accommodate the trends/patterns within the data, including both the increasing frequency over time as well as the periodic jumps to lower frequency regimes. The residual plot depicts lower magnitudes as well when compared to Model MMPP-3 (3). However, Table \ref{table:residualstats} provides strong evidence that there still exists auto-correlation within the residuals, indicating the presence of data features not taken into account by the Markov modulation. Further, the high number of regimes, while attempting to compensate for the increasing trend, unfortunately detract from understanding of the model. Ultimately, this is an issue with the structure of the MMPP model. A more natural approach may be to account for this trend elsewhere instead of using discretised regimes. This removes the dominating effect of the trend and ideally, reduces the number of regimes. It may be difficult to extract explainable information from a model with a very high number of regimes, as evidenced by the MMPP prediction plot above (which we note resembles the regime filtering plot for the most likely regime per day). Aside from the obvious upwards trend, it requires significant effort to develop additional insight into the causes of the hidden regimes. 

Finally, the 3-state MMNPP residuals resemble white-noise. This model serves as a useful composite model that combines the `non-parametric' fit of the MMPP with the model interpretability provided by the non-homogeneous Poisson model. The latter component  allows for the tractable incorporation of domain knowledge whilst maintaining superior goodness-of-fit. The number of regimes in the MMNPP is only 3, which suitably caters for further diagnosis and model improvement. Also, by adding structural features through the ODP GLM, we have actually improved marginally on the fit of the MMPP, as can be seen in the residual statistics shown in Table \ref{table:residualstats}. This has also removed the auto-correlation present in the NHPP and MMPP residuals. In summary, this case study demonstrates that the MMNPP produces a significant number of advantages through its extensions to the NHPP and MMPP models by allowing for expert experience and features extracted through data analysis to be integrated into the modelling process, while still producing excellent fits to data and accounting for characteristics such as auto-correlation.

\begin{remark}
	We remark that the MMNPP provides clear advantages for out-of-sample comparisons. The MMNPP model overcomes a key weakness of the original MMPP model in terms of out-of-sample predictions, in that MMPP predictions are (without adjustment) unable to maintain the time-structure of regimes that are extracted from the data set. In other words, the MMNPP allows for known effects which do not follow Markov arrivals to be modelled by the non-homogeneous component. Further, we have already commented that the structure of the MMPP seems unsuitable to capture the upwards trend displayed in the data and it is expected that out-of-sample comparisons will heavily favour the MMNPP model. For example, if this trend were continue in the future, the MMPP model would be unable to account for the higher frequency. Thus, we do not provide numerical results of the out-of-sample comparison here.
\end{remark}

\section{Conclusion}\label{sec:Conc}
In this paper, a Markov-modulated non-homogeneous Poisson process framework was introduced to model counting processes. We produced an extension of the standard MMPP models through a flexible frequency perturbation measure, which allows for the inclusion of both non-recurring and periodic data features to be readily incorporated into analysis. The proposed extension is beneficial as it not only allows known factors of interest to be included in a tractable manner but also provides some control over the underlying drivers of the observed events that are modelled through the hidden Markov chain. This produces a more comprehensive, accurate and realistic count model compared to the original MMPP approach. This method is also intuitive as the MMNPP is shown to be an operational time transform of the original MMPP. An efficient EM calibration procedure is derived, allowing the model to be applied in cases where there are large numbers of observations. In these situations, issues of computational cost and numerical stability are of great concern. These barriers to practical implementation are addressed through various computational and theoretical innovations such as a scaling procedure for the backwards-forwards recursions utilised in the calibration algorithm and methods to improve calibration speeds. 
Our theoretical results and proposed implementation procedure are illustrated through a case study using real insurance data and insights that are provided through the MMNPP model are discussed. The procedure is shown to generate interpretable regimes in the empirical study and the MMNPP analysis framework is shown to generate additional results that can allow modellers to delve more deeply into the data if it is deemed necessary. Comparisons to alternative models demonstrate the advantages of the MMNPP model as an attractive composite model that provides both superior goodness-of-fit as well as significant model interpretability and tractability.

%
%
%



\section*{Acknowledgements}
The authors are grateful to two anonymous referees for detailed comments, which led to substantial improvements to the paper. Earlier versions of this paper were presented at the 8th and 9th \emph{Australasian Actuarial Education and Research Symposium} in Sydney (Australia), the \emph{22nd International Congress on Insurance: Mathematics and Economics} in Sydney (Australia), the \emph{Actuarial Risk Modelling and Extreme Values Workshop} in Canberra (Australia),  the 4th \emph{European Actuarial Journal Conference} in Leuven (Belgium), the ASTIN Colloquium in Cape Town (South Africa) as well as at seminars held at the universities of Lyon and Copenhagen in 2019.  Some results in this paper were also presented at the Australian Actuaries Institute General Insurance Seminar in November 2018 (and awarded the Taylor Fry Silver Award); see \citet{AvTaWoXi18}. The authors are grateful for constructive comments received from colleagues who attended these events.

This research was  supported under Australian Research Council's Linkage (LP130100723, with funding partners Allianz Australia Insurance Ltd, Insurance Australia Group Ltd, and Suncorp Metway Ltd) and Discovery (DP200101859) Projects funding schemes. Furthermore, Alan Xian acknowledges financial support by the Australian Government Research Training program, as well the UNSW Business School through supplementary scholarships. The views expressed herein are those of the authors and are not necessarily those of the supporting organisations.



\section*{Data and Code}
We are unable provide the dataset that was used in the empirical case study due to confidentiality. However, simulated data sets with similar features, as well as all relevant codes, can be found at \url{https://github.com/agi-lab/reserving-MMNPP}.

\section*{References}

\bibliographystyle{elsarticle-harv}

\bibliography{biblio}

\newpage

\renewcommand{\theHsection}{A\arabic{section}}
\appendix

\section{Appendix: Proof of Lemmas \ref{lem:MMNPPmarkov} and \ref{lem:MMNPPiffcond}}

	\begin{pfof}{Lemma \ref{lem:MMNPPmarkov}}\label{app:lem1}
	For some constants $s,t$, it is known that
	\begin{align*}
	\mathbb{P} \left[ N_M(\rho^{-1}(t))|N_M(\rho^{-1}(i)); 0 \le i < s \right] & = \mathbb{P}\left[N(t)|N(i); 0 \le i < s \right] \\
	& = \mathbb{P}\left[N(t)|N(s)\right] \\ 
	& = \mathbb{P}\left[N_M(\rho^{-1}(t))|N_M(\rho^{-1}(t)) \right]
	\end{align*}
	Thus, the new process $\{M(t),N_M(\rho^{-1}(t))\}$ retains the Markov property.
	\end{pfof}

	\begin{pfof}{Lemma \ref{lem:MMNPPiffcond}}\label{app:lem2}
	$(\implies)$ Beginning from the definition of a homogeneous MMPP,
	\begin{align*}
	\mathbb{P} \left[ N_M(t_1) = n, N_M(t_2) =n \right] & = \exp\left[ -\int_{t_1}^{t_2} \lambda_{M(x)} dx \right] \\ 
	& = \exp \left[ - \sum_{k=1}^{m} \int_{u_{k-1}}^{u_k} \lambda_{s_k} dx \right] \\ 
	& = \exp \left[ -\sum_{k=1}^{m} \lambda_{s_k} (u_{k} - u_{k-1}) \right]
	\end{align*}
	Here, $s_k$ represents the regime of the process $M(t)$ in the interval $I_k$.
	
	$(\impliedby)$ Beginning from the expression,
	\begin{align*}
	\lim_{h \rightarrow 0} \frac{1- \mathbb{P} \left[ N_M(t) = n, N_M(t+h) =n \right]}{h} & = \lim_{h \rightarrow 0} \frac{1 - \exp \left[ -\sum_{k=1}^{m} \lambda_{s_k} (u_{k} - u_{k-1}) \right]}{h}
	\end{align*}
	for the regime intervals that fall between times $t$ and $t+h$. Note that as $h$ approaches zero, the right expression eventually becomes a time interval within a single regime period, so that
	\begin{align*}
	\lim_{h \rightarrow 0} \frac{1 - \exp \left[ -\sum_{k=1}^{m} \lambda_{s_k} (u_{k} - u_{k-1}) \right]}{h} & = \lim_{h \rightarrow 0} \frac{1 - \exp \left[ - \lambda_{M(t)} (t - (t+h)) \right]}{h} = \lambda_{M(t)}.
	\end{align*}
	Thus, homogeneity is achieved as $\lambda_{M(t)}$ is the constant claim intensity at time $t$ for state of the process $M$ at time $t$.
	\end{pfof}

\section{Appendix: Proof of Proposition \ref{prop:addMMNPP}}

\begin{pfof}{Proposition \ref{prop:addMMNPP}}\label{app:prop1}
	This proposition is a very intuitive result. As the Markov process $M(t)$ is the same in both MMPP processes, it suffices to prove that the combined process $N_{M}^{1,2} (t)$ is a conditional Poisson process with intensity $\lambda_{M}(t) = \lambda^1_{M}(t) + \lambda^2_{M}(t)$ in order to obtain the result that $\{M(t),N^{1,2}_{M} (t)\}$ is the required MMPP. This follows directly from the definition of Poisson processes. We begin by noting that each $N^i_{M}(t)$ is a conditional Poisson process, i.e., $N^i_{M}(t)$ for $i=1,2$ has the following four properties:
	\begin{enumerate}
		\item $N^{i}_{M}(0) = 0$;
		\item $N^i_{M}(t)$ has independent increments;
		\item $\mathbb{P} [N^i_M (t+\Delta t) - N^i_M(t) = 1] = \lambda^i_M (t) \Delta t + o(\Delta t)$; and
		\item $\mathbb{P} [N^i_M (t+\Delta t)-N^i_M (t) \ge 2] = o(\Delta t)$,
	\end{enumerate}
	where $o(\Delta t)$ is little-o notation for $\lim_{\Delta t\rightarrow 0} \frac{o(\Delta t)}{\Delta t} =0$. We wish to prove that these properties also hold for $N_{M}^{1,2} (t)$.
	
	The first property is easily seen for $N_M^{1,2} (t)$ as $N^{1,2}_M(0) = N^{1}_M(0) + N^{2}_M(0) = 0 + 0 = 0$. The second property holds by using that the conditional Poisson processes $N^i_M (t)$ are independent for $i = 1,2$ and that each process individually has independent increments. Thus, $N^i_M(t)$ must also have independent increments.
	
	For the third property, we have that
	\begin{align*}
	\mathbb{P} [N_M^{1,2} (t+\Delta t) - N_M^{1,2}(t) = 1] & = \mathbb{P} [N_M^1(t+\Delta t) - N_M^1(t) = 1, N_M^2(t+\Delta t) 		- N_M^2(t) = 0] \\
	& + \mathbb{P} [N_M^1(t+\Delta t) - N_M^1(t) = 0, N_M^2(t+\Delta t) - N_M^2(t)=1] \\
	& = (\lambda^1_{M(t)}(t)\Delta t + o(\Delta t))(1-\lambda^2_{M(t)}(t)\Delta t + o(\Delta t)) \\
	& + (1 - \lambda^1_{M(t)}(t)\Delta t + o(\Delta t))(\lambda^2_{M(t)}(t)\Delta t + o(\Delta t)) \\
	& \hspace{1em} \text{due to the independence between the two processes} \\
	& = [\lambda^1_{M(t)}(t) + \lambda^2_{M(t)}(t)]\Delta t + o(\Delta t).
	\end{align*}
	
	The final property can be shown in a similar manner. Thus, the combined process $N_M^{1,2}(t)$ is a conditional Poisson process with a claim intensity given by $\lambda^1_{M(t)}(t) + \lambda^2_{M(t)}(t)$.
\end{pfof}

\section{Appendix: Derivations for EM algorithm estimators}

\subsection{The estimator for the number of regime changes: $\boldsymbol{\hat{a}_{i,j}}$}
Beginning with the estimator used for the number of regime/state changes, $\hat{a}_{i,j}$, the integrand in Equation \eqref{eq:miestimator} is
\begin{align*}
& \mathbb{P} [M(s^-)=i,M(s)=j | N_M(\rho^{-1}(t)), 0 \le t \le T] \nonumber \\
& = \frac{\mathbb{P} [M(s)=j|M(s^-)=i] \mathbb{P} [M(s^-)=i] \mathbb{P} [N_M(\rho^{-1}(t)), 0 \le t \le T | M(s^-)=i, M(s)=j]}{\mathbb{P} [N_M(\rho^{-1}(t)), 0 \le t \le T]}.
\end{align*}
The initial term of the numerator is simply $q_{i,j}$ and by using the conditional independence, the last term of the numerator is decomposed to
\begin{align*}
& \mathbb{P} [N_M(\rho^{-1}(t)), 0 \le t \le T | M(s^-)=i, M(s)=j] \\
& = \frac{\mathbb{P} [N_M(\rho^{-1}(t)), 0 \le t < s, M(s^-)=i]}{\mathbb{P} [M(s^-)=i]} \mathbb{P} [N_M(\rho^{-1}(t)), s \le t \le T | M(s)=j].
\end{align*}
Thus, we obtain
\begin{align}
\hat{a}_{i,j} 	& = \frac{q_{i,j}}{\mathbb{P} [N_M(\rho^{-1}(t)), 0 \le t \le T]} \nonumber \\
& \times \int_{0}^{T} \mathbb{P} [N_M(\rho^{-1}(t)), 0 \le t < s, M(s^-)=i] \mathbb{P} [N_M(\rho^{-1}(t)), s \le t \le T | M(s)=j] ds 
\end{align}
Each of these probabilities can be rewritten in terms of the transition densities from the previous section. The probability $\mathbb{P} [N_M(\rho^{-1}(t)), 0 \le t \le T]$ is simply the likelihood of the observed claim arrivals which can be expressed in matrix form as
\begin{equation}
\mathbb{P} [N_M(\rho^{-1}(t)), 0 \le t \le T] = \boldsymbol{\pi}(Q^0,\Lambda^0 ) \bigg[ \prod_{k=1}^{n} \boldsymbol{f}^{\delta_k}(t_{k-1},t_k - t_{k-1}) \bigg] \boldsymbol{1}
\end{equation}
where $\boldsymbol{\pi}(Q^0,\Lambda^0 )$ is the starting distribution of state probabilities given the initial parameter estimates $Q^0$ and $\Lambda^0$, $\boldsymbol{1}$ is a $r\times 1$ vector of ones and $t_0 = 0$. Also, $k$ is now a counter for both event types and $n$ is the total number of original events and exposure changes. In a similar manner, the probabilities within the integrand are
\begin{align}
\mathbb{P} [N_M(\rho^{-1}(t)), 0 \le t < s, M(s^-)=i] & = \boldsymbol{\pi}(Q^0,\Lambda^0 ) \bigg[ \prod_{k=1}^{N_M(\rho^{-1}(s)^-)} \boldsymbol{f}^{\delta_k}(t_k - t_{k-1}, t_{k-1}) \bigg] \boldsymbol{\bar{F}} (t_{N_M(\rho^{-1}(s)^-)}, s - t_{N_M(\rho^{-1}(s)^-)}) \boldsymbol{1}_i, \\
\mathbb{P} [N_M(\rho^{-1}(t)), s \le t \le T | M(s)=j] & = \boldsymbol{1}^\intercal_{j} \boldsymbol{f}^{\delta_{N_M(\rho^{-1}(s)^-)+1}} (s, t_{N_M(\rho^{-1}(s)^-)+1} - s) \bigg[ \prod_{k=N_M(\rho^{-1}(s)^-)+2}^{n} \boldsymbol{f}^{\delta_k}(t_{k-1},t_k - t_{k-1}) \bigg] \boldsymbol{1},
\end{align}
where $\boldsymbol{1}_i$ is a vector of zeroes except for the $i$-th entry which is one. Thus, the final expression for the estimator of $a_{i,j}$ is
\begin{align}
\hat{a}_{i,j} 	& = \frac{q^0_{i,j}}{\boldsymbol{\pi}(Q^0,\Lambda^0) \bigg[ \prod_{k=1}^{n} \boldsymbol{f}^{\delta_k}(t_{k-1},t_k - t_{k-1}) \bigg] \boldsymbol{1}} \nonumber \\
& \times \Bigg[\int_{0}^{T} \boldsymbol{\pi}(Q^0,\Lambda^0) \bigg[ \prod_{k=1}^{N_M(\rho^{-1}(s)^-)} \boldsymbol{f}^{\delta_k}(t_{k-1},t_k - t_{k-1}) \bigg] \boldsymbol{\bar{F}} (t_{N_M(\rho^{-1}(s)^-)}, s - t_{N_M(\rho^{-1}(s)^-)}) \boldsymbol{1}_i \nonumber \\ 
& \times \boldsymbol{1}^\intercal_{j}  \boldsymbol{f}^{\delta_{N_M(\rho^{-1}(s)^-)+1}}(s,t_{N_M(\rho^{-1}(s)^-)+1} - s) \bigg[ \prod_{k=N_M(\rho^{-1}(s)^-)+2}^{n} \boldsymbol{f}^{\delta_k}(t_{k-1},t_k - t_{k-1}) \bigg] \boldsymbol{1} ds \Bigg] .
\end{align}

\subsubsection{The estimator for the number of events of interest in each regime: $\boldsymbol{\hat{n}_{i}}$}

Continuing from Equation \eqref{eq:niestimator}, we obtain
\begin{align}
\hat{n}_{i}		& = \mathbb{E}[n_{i}|N_M(\rho^{-1}(t)), 0 \le t \le T] \nonumber \\
& = \sum_{k=1}^{n} \mathbb{P} [M(t_{k}) = i | N_M(\rho^{-1}(t)), 0 \le t \le T] \nonumber \\
& = \sum_{k=1}^{n} \frac{\mathbb{P} [M(t_{k}) = i,N_M(\rho^{-1}(t)), 0 \le t \le T]}{\mathbb{P} [N_M(\rho^{-1}(t)), 0 \le t \le T]} \nonumber\\
& = \frac{1}{\boldsymbol{\pi}(Q^0,\Lambda^0) \bigg[ \prod_{k=1}^{n} \boldsymbol{f}^{\delta_k}(t_{k-1},t_k - t_{k-1}) \bigg] \boldsymbol{1}} \nonumber \\
& \times \sum_{k=1}^{n} \boldsymbol{\pi}(Q^0,\Lambda^0) \bigg[ \prod_{l=1}^{N_M(\rho^{-1}(t_k))} \boldsymbol{f}^{\delta_l}(t_{l-1},t_l - t_{l-1}) \bigg] \boldsymbol{1}_i \boldsymbol{1}_i^\intercal \bigg[ \prod_{l=N_M(\rho^{-1}(t_k)+1}^{n} \boldsymbol{f}^{\delta_l}(t_{l-1},t_l - t_{l-1}) \bigg] \boldsymbol{1}.
\end{align}

\subsubsection{The estimator for the time spent in each regime: $\boldsymbol{\hat{T}_{i}}$}
Using the expression for $\hat{T}_i$ from Equation \eqref{eq:tiestimator} and following a similar procedure to the $\hat{m}_{i,j}$ calculations, it can be seen that
\begin{align}
\hat{T}_i 	& = \int_{0}^{T} \mathbb{P} [M(s) = i | N_M(\rho^{-1}(t)), 0 \le t \le T ] ds \nonumber\\
& = \frac{1}{\boldsymbol{\pi}(Q^0,\Lambda^0 ) \bigg[ \prod_{k=1}^{n} \boldsymbol{f}^{\delta_k}(t_{k-1},t_k - t_{k-1}) \bigg] \boldsymbol{1}} \nonumber \\
& \times  \int_{0}^{T} \mathbb{P} [M(s) = i, N_M(\rho^{-1}(t));0\le t < s] \mathbb{P} [N_M(\rho^{-1}(t)); s \le t \le T | M(s) = i] ds \nonumber \\
& = \frac{1}{\boldsymbol{\pi}(Q^0,\Lambda^0) \bigg[ \prod_{k=1}^{n} \boldsymbol{f}^{\delta_k}(t_{k-1},t_k - t_{k-1}) \bigg] \boldsymbol{1}} \nonumber \\
& \times \Bigg[\int_{0}^{T} \boldsymbol{\pi}(Q^0,\Lambda^0 ) \bigg[ \prod_{l=1}^{N_M(\rho^{-1}(s)^-)} \boldsymbol{f}^{\delta_l}(t_{l-1},t_l - t_{l-1}) \bigg] \boldsymbol{\bar{F}} (t_{N_M(\rho^{-1}(s)^-)},s - t_{N_M(\rho^{-1}(s)^-)}) \boldsymbol{1}_i \nonumber \\ 
& \times \boldsymbol{1}^\intercal_{j}  \boldsymbol{f}^{\delta_{N_M(\rho^{-1}(s)^-)+1}}(s,t_{N_M(\rho^{-1}(s)^-)+1} - s) \bigg[ \prod_{l=N_M(\rho^{-1}(s)^-)+2}^{n} \boldsymbol{f}^{\delta_l}(t_{l-1},t_l - t_{l-1}) \bigg] \boldsymbol{1} ds \Bigg].
\end{align}
In the above, the event $\{M(s^-)=i\}$ has been replaced with $\{M(s)=i\}$ which does not change the density due to the fact that $\{M(s)\}$ is continuous in probability.

\subsubsection{The estimator for the operational time spent in each regime: $\boldsymbol{\hat{T}^*_{i}}$}
The results from the previous section are easily adapted to obtain the following estimator for $T^*_{i}$:
\begin{align}
\hat{T}_i^* 	& = \int_{0}^{T} \mathbb{P} [M(s) = i | N_M(\rho^{-1}(t)), 0 \le t \le T ] \gamma(s) ds \nonumber\\
& = \frac{1}{\boldsymbol{\pi}(Q^0,\Lambda^0) \bigg[ \prod_{k=1}^{n} \boldsymbol{f}^{\delta_k}(t_{k-1},t_k - t_{k-1}) \bigg] \boldsymbol{1}} \nonumber \\
& \times \Bigg[\int_{0}^{T} \boldsymbol{\pi}(Q^0,\Lambda^0) \bigg[ \prod_{l=1}^{N_M(\rho^{-1}(s)^-)} \boldsymbol{f}^{\delta_l}(t_{l-1},t_l - t_{l-1}) \bigg] \boldsymbol{\bar{F}} (t_{N_M(\rho^{-1}(s)^-)}, s - t_{N_M(\rho^{-1}(s)^-)}) \boldsymbol{1}_i \nonumber \\ 
& \times \boldsymbol{1}^\intercal_{j}  \boldsymbol{f}^{\delta_{N_M(\rho^{-1}(s)^-)+1}} (s,t_{N_M(\rho^{-1}(s)^-)+1} - s) \bigg[ \prod_{l=N_M(\rho^{-1}(s)^-)+2}^{n} \boldsymbol{f}^{\delta_l}(t_{l-1},t_l - t_{l-1}) \bigg] \boldsymbol{1} \gamma(s) ds \Bigg]
\end{align}
\end{document}